\documentclass[preprint,journal]{vgtc}            

\onlineid{2033}

\vgtccategory{Research}

\title{ESVR: 3D Ellipsoid-based Sparse Volume Rendering via Structure-aware Primitive Learning and Per-primitive Ray Sampling}

\author{%
  \authororcid{Suemin Jeon\textsuperscript{*}}{0009-0007-1013-4845},
  \authororcid{Youjin Kim\textsuperscript{*}}{0009-0005-1978-0400},
  \authororcid{Jungwoo Park}{0009-0008-5728-6667},
\authororcid{Kyungryun Lee}{0009-0006-3893-4177},
  and
   \authororcid{Won-Ki Jeong}{0000-0002-9393-6451}
}

\authorfooter{
  \item
  	Suemin Jeon, Youjin Kim, Jungwoo Park, Kyungryun Lee and Won-Ki Jeong are with Korea University.
  	E-mail: {orangeblush, zinzinyou, oscar6090, krlee0000, wkjeong}@korea.ac.kr
   \item \textsuperscript{*}~These authors contributed equally to this work.
}


\graphicspath{{figs/}{figures/}{pictures/}{images/}{./}} 

\usepackage{tabu}                      
\usepackage{booktabs}                  
\usepackage{lipsum}                    
\usepackage{mwe}                       
\usepackage{ccicons}                   
\usepackage{amsmath}
\usepackage{algorithm}
\usepackage{algpseudocode}
\usepackage{amssymb}
\usepackage{makecell}
\usepackage{color}
\usepackage{multirow} 
\usepackage{booktabs}
\usepackage{tabularx}
\usepackage{xcolor}
\usepackage{makecell}
\usepackage{multirow}
\usepackage{gensymb}

\usepackage{mathptmx}                  

\newcommand{\sm}[1]{{\color{black} #1}} 

\begin{document}



\teaser{
  \centering
  \includegraphics[width=\linewidth]{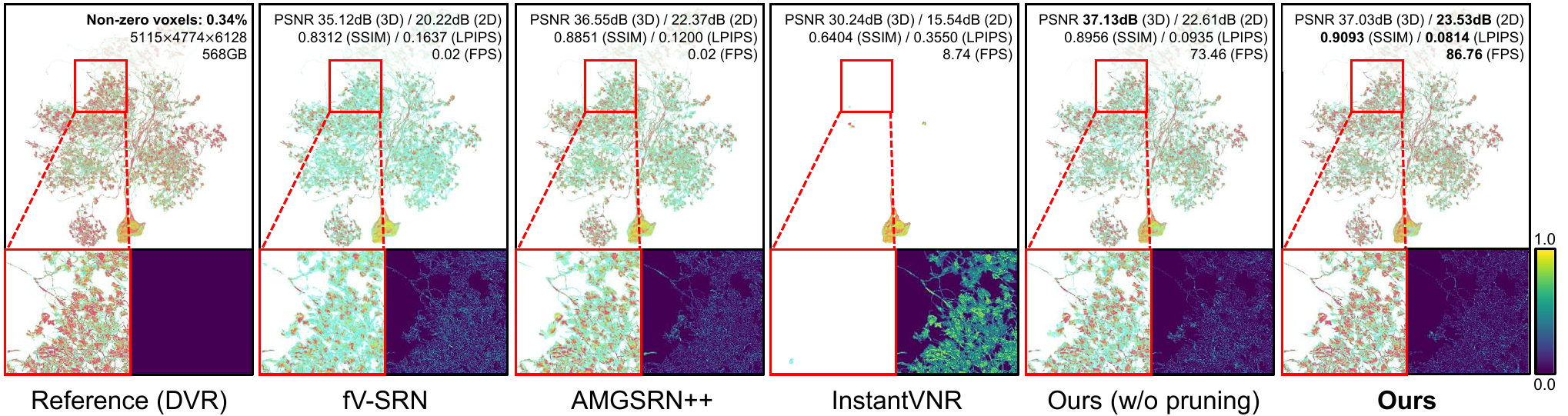}
  \vspace{-0.27in}
  \caption{%
  	Rendering results on the Hemibrain (lLN2P) dataset under a model-size-matched setting (5.8MB).
    Our method reconstructs detailed sparse structures more accurately while maintaining real-time framerates. Top-right text shows each method's 3D PSNR, 2D PSNR, SSIM, LPIPS, and FPS. The bottom row shows a zoomed-in view and the corresponding 2D \sm{absolute per-pixel} error map.
  }
  \label{fig:teaser}
}

\abstract{%
Efficient representation and rendering of large-scale sparse volumetric data remain challenging in scientific visualization, as meaningful structures often occupy only a small fraction of the spatial domain.
While direct volume rendering (DVR) provides high-quality visualization, its computational and memory costs scale poorly with data size.
Recent advances in 3D Gaussian Splatting (3DGS) address this challenge by representing volumetric scenes with compact geometric primitives, enabling efficient, high-fidelity rendering.
However, existing 3DGS-based methods learn from DVR rendered images rather than raw volumes, leading to information loss and limiting flexible transfer function control for interactive exploration.
To address these limitations, we propose ESVR, an ellipsoid-based sparse volume rendering framework that directly learns and renders volumetric data in 3D space. 
Our method combines differentiable ellipsoidal primitives with bounded support, structure-aware primitive learning with complementary pruning, and a per-primitive ray sampling strategy for fast and accurate transfer function mapping. 
To support large-scale datasets, we further introduce a chunk-based optimization scheme with ghost ellipsoids, providing boundary context during training.
Across large sparse datasets, ESVR achieves up to four orders of magnitude compression and real-time rendering at 43--223 FPS while maintaining competitive reconstruction quality.
}

\maketitle

\section{Introduction}

In scientific visualization, efficiently representing and rendering large-scale sparse volumetric data remains a long-standing challenge~\cite{ganter2019analysis,kim2024neuralvdb,sicat2014sparse,zellmann2019linear,zellmann2019hybrid,kahler2003interactive,labschutz2015jittree, vega2005high}. 
%
%
In many real-world datasets, such as vascular structures in medical images and neuronal structures in microscopy images, meaningful signals occupy only a small fraction of the spatial domain, while most voxels are empty or homogeneous (see Fig.~\ref{fig:teaser}, only 0.34\% of the volume contains nonzero values). 
As a result, conventional direct volume rendering (DVR)~\cite{levoy2002display, ljung2016state} incurs high computational and memory overhead due to dense sampling across uninformative regions. 
%
Prior work addresses this inefficiency using specialized data structures and empty-space skipping~\cite{museth2013openvdb,hadwiger2017sparseleap,zellmann2019binned,wald2021faster}.
However, these approaches remain tied to grid representations, making them resolution-dependent and limiting adaptability to 
underlying data structures.
%
%

%
%

Recently, learning-based scene representation methods have been widely adopted in volume visualization (VolVis) as an alternative to conventional grid-based representations. 
One of them is Scene Representation Networks (SRNs)~\cite{lu2021compressive, wurster2022deep, weiss2022fast, han2022coordnet, muller2022instant, wu2023interactive, wurster2023adaptively}, which compress the original volume into compact neural representations. 
While these approaches reduce memory and eliminate direct access to raw volumes during rendering, their dense per-point evaluation introduces substantial computational cost, hindering real-time performance.
Another learning-based method, 3D Gaussian Splatting (3DGS)~\cite{kerbl20233d}, has shown strong performance in scene representation by using spatially adaptive primitives and efficient hardware rasterization. 
%
These properties, along with resolution independence and elimination of repeated network inference, make 3DGS promising for VolVis~\cite{niedermayr2024application, kleinbeck2025multi, tang2025ivr, tang2025texgs, ai2025nli4volvis}.

\begin{figure}[t]
 \centering
  \includegraphics[width = 0.88\linewidth]{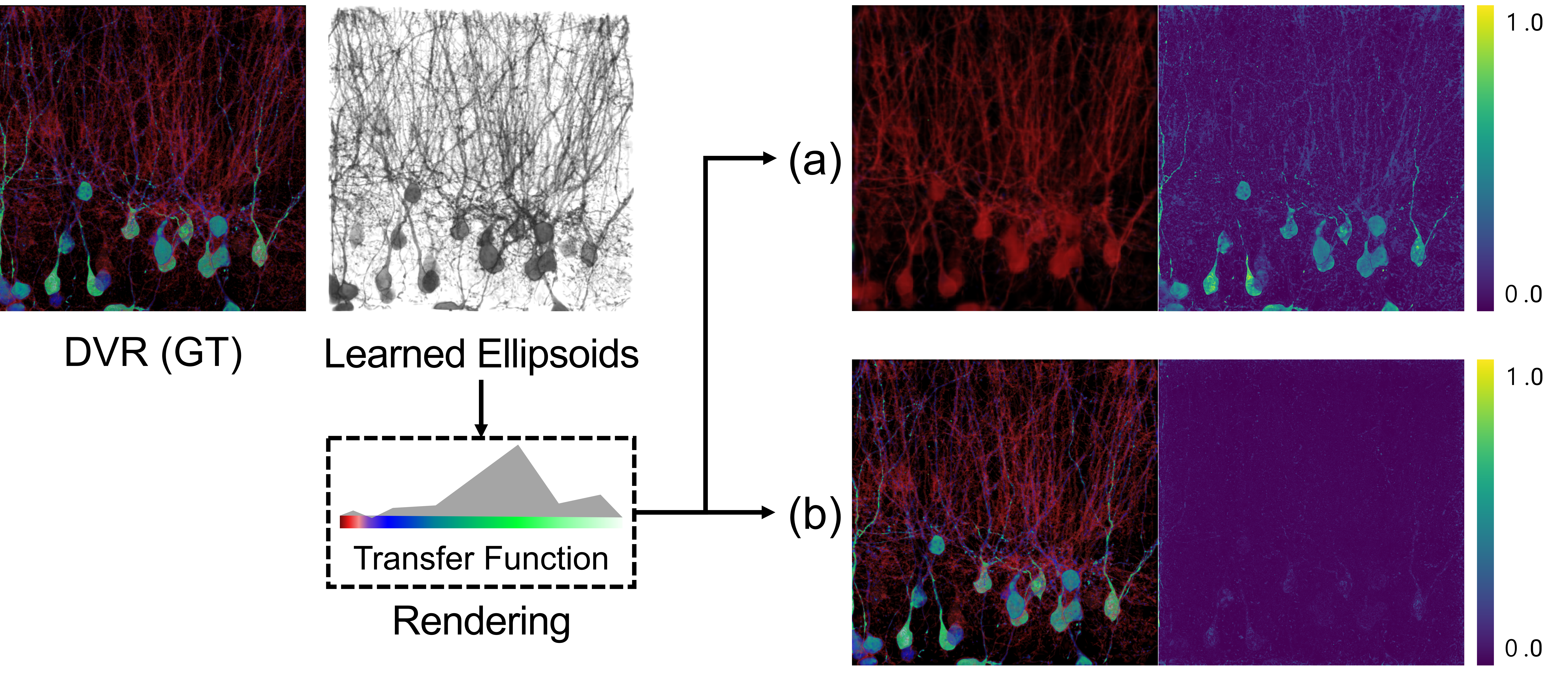}
\vspace{-0.13in}
   \caption{Rendering results on the neuron dataset after TF application, 
   comparing conventional (a) 2D rasterization in 3DGS with (b) our per-primitive ray sampling. The errormap shows \sm{absolute per-pixel} errors.
   }
 \label{fig:tf_limitation}
\end{figure}

The motivation of this work arises from the observation that Gaussian primitives are well-suited for representing sparse volumetric data, as they can be placed adaptively only in regions containing meaningful signal, enabling efficient storage and rendering.
%
However, current 3DGS-based volume visualization methods face several limitations: 
First, existing methods rely on image-level supervision from DVR-rendered images generated with fixed transfer functions (TFs), which inevitably causes information loss and limits interactive exploration of the raw input volume. 
Several works address this limitation by combining multiple Gaussian fields constructed with different TFs~\cite{tang2025ivr,tang2025texgs}. 
However, this strategy incurs additional memory overhead and depends on predefined scene context. 
Second, the original 3DGS relies on alpha blending of 2D rasterized Gaussians under a pre-classification scheme, in which each Gaussian is assigned a color prior to compositing. 
This formulation is not aligned with TF mapping in alpha-blended ray casting, leading to inaccurate TF evaluation and rendering artifacts, particularly in regions with overlapping primitives (Fig.~\ref{fig:tf_limitation}).
Third, conventional 3DGS employs heuristic adaptive density control (ADC) based on 2D projections, which fails to capture local structural characteristics of the input volume. 
As a result, 3DGS often generates an excessive overlapping Gaussians, leading to artifacts and increased memory consumption. 
Moreover, the long-tailed spatial support of Gaussians further exacerbates inter-primitive overlap and redundancy.

To address these limitations, we propose ESVR, an \textbf{E}llipsoid-based  \textbf{S}parse  \textbf{V}olume  \textbf{R}endering framework that directly learns and renders volumetric data in 3D space. 
ESVR combines the memory efficiency of primitive-based scene representations with rendering quality comparable to DVR, making it particularly well suited for sparse volume rendering. 
Instead of Gaussian primitives, ESVR employs differentiable ellipsoidal primitives with compact spatial support, reducing inter-primitive overlap and the resulting artifacts.
Building on this representation, ESVR incorporates a structure-aware primitive learning framework: ADC ensures sufficient coverage of the input volume, while complementary structure-aware pruning removes redundant primitives without sacrificing structural anisotropy or fine details.
In addition, ESVR performs GPU-accelerated per-primitive ray sampling directly in 3D space, enabling fast and accurate TF evaluation without screen-space approximations as in 3DGS. 
To support large-scale datasets, ESVR further incorporates a chunk-based optimization scheme with ghost ellipsoids, preserving boundary context during training and enabling efficient optimization of volumes exceeding single-GPU memory.
The following summarizes our main contributions:
%
\begin{itemize}
    \item We propose a novel ellipsoid-based 3D volume rendering framework that represents and renders volumetric data using adaptive differentiable ellipsoidal primitives. By directly fitting compact-support ellipsoids to the input volume, the framework yields an efficient yet expressive representation for sparse volumetric data. 

    \item We present a structure-aware primitive learning framework that combines ADC with complementary pruning for compact and efficient scene representations. We further extend this with a scalable chunk-based training scheme using ghost ellipsoids, enabling large-volume optimization without boundary artifacts.

    \item We develop a GPU-accelerated per-primitive ray sampling strategy that evaluates 3D ellipsoids without 2D hardware rasterization, enabling accurate TF mapping for alpha blending along rays while achieving efficient rendering.

\end{itemize}

\vspace{-0.17in}
\begin{figure*}[!t]
 \centering
 \includegraphics[width=1.0\textwidth]{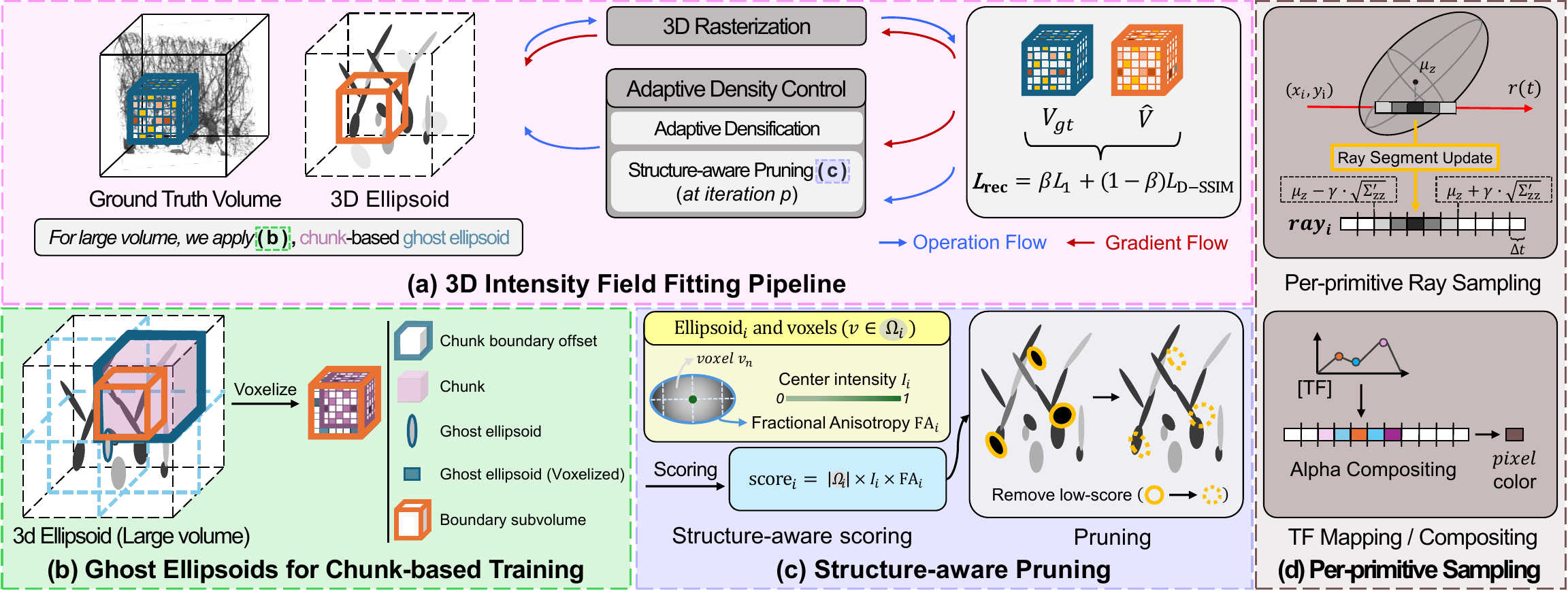}
 \vspace{-0.23in}
   \caption{
   Overview of our proposed framework: (a) 3D intensity field fitting pipeline, (b) chunk-based large volume training using ghost ellipsoid and (c) structure-aware pruning strategy, (d) per-primitive sampling and transfer function mapping of the intensity-fitted ellipsoids. }
 \label{fig:pipeline}
\end{figure*}
\section{Related Work} 
\vspace{-0.05in}
\subsection{Primitive-based Volume Rendering}
\label{sec:rel-primitive}
Primitive-based volume rendering approximates volumetric fields by splatting kernel primitives onto the image plane~\cite{zwicker2001ewa, chen2004hardware, vega2005high, neophytou2006gpu}. 
Recent advances have extended this paradigm with 3DGS~\cite{kerbl20233d}, enabling efficient rendering via spatially adaptive primitives and hardware-accelerated rasterization.
%
%
Several works have applied 3DGS to volume visualization~\cite{tang2025ivr, ai2025nli4volvis, tang2025texgs, yao2025volseggs, niedermayr2024application, kleinbeck2025multi}, with improved rendering speed and compactness.
%
However, most approaches optimize primitives from DVR-rendered images under predefined TFs, tying representations to specific TF configurations and limiting generalization to arbitrary TF manipulation.
%
%
Furthermore, 3DGS relies on screen space rasterization, which is not well-aligned with volumetric signal reconstruction. 
This is particularly evident when primitives overlap, as compositing 2D rasterized Gaussians in screen space cannot correctly recover the underlying 3D intensity distribution.
%
%
Recent efforts have partially addressed these limitations. 
iVR-GS~\cite{tang2025ivr} trains separate models across multiple TF configurations and composites them, but supports only palette-based adjustments. \sm{VEG~\cite{dyken2026volume}} learns a scalar field from training images to enable unseen TF application, yet remains sensitive to training TF coverage and quality.
\sm{Talegaonkar et al.~\cite{talegaonkar2025volumetrically} reformulate Gaussian rasterization to align with the volumetric ray integration model, and $R^2$-Gaussian~\cite{cai2024radiative} incorporates a pseudo-volume for volumetric consistency; however, both remain image-supervised without direct optimization against ground-truth volumes.}
%
%
In contrast, our method learns from the volumetric intensity field and performs rendering in 3D space, avoiding image-based supervision and screen-space aggregation.

%
%


\vspace{-0.095in}
\subsection{Neural Representation for Large Sparse Volume}
%
Neural representations model volumetric data as continuous functions, offering an alternative to raw volumetric grids.
Early approaches, commonly referred to as SRNs~\cite{lu2021compressive,wurster2022deep,weiss2022fast,han2022coordnet,muller2022instant,wu2023interactive,wurster2023adaptively}, enable compact representations and support downstream tasks such as temporal and spatial super-resolution~\cite{han2022coordnet,tang2024stsr}.
However, these methods are computationally expensive due to dense sampling and per-query inference~\cite{wurster2022deep,lu2021compressive,han2022coordnet}.
%
%
%
%
%
Subsequent improvements such as latent grids~\cite{weiss2022fast} and hash encodings~\cite{muller2022instant} significantly improved efficiency.
To further address scalability, Wu et al.~\cite{wu2023interactive} introduce an out-of-core sampling strategy combined with macro-cell acceleration to enable interactive exploration of large volumes.
Wurster et al.~\cite{wurster2023adaptively,wurster2025amgsrn++} propose adaptive grid-based representations that improve efficiency for large sparse volumes.
These approaches are well suited for continuous or smoothly varying volumetric signals.
%
%
%
Despite these advances, such methods remain expensive for large-scale sparse datasets~\cite{wurster2022deep,lu2021compressive,han2022coordnet,wurster2025amgsrn++},
and less efficient when meaningful structures occupy only a small fraction of the spatial domain.
In contrast, our method explicitly places primitives only in signal-bearing regions and retrieves information through per-primitive sampling, avoiding redundant memory access and improving efficiency for sparse volumetric data.
\vspace{-0.095in}
\subsection{Improving Memory and Rendering Efficiency 
in 3DGS}
%
%
%
Several works improve the efficiency of 3DGS by reducing primitive count and accelerating rendering. 
Methods such as Speedy-Splat~\cite{hanson2025speedy} optimize rasterization, \sm{while compression techniques reduce the memory footprint via quantization and compact attribute encoding~\cite{niedermayr2024compressed, papantonakis2024reducing}}.
%
EVER~\cite{mai2024ever} introduces compact constant-density ellipsoidal primitives that better capture surface and edge structures, motivating our use of ellipsoids. 
However, its hard-boundary formulation is not suited to optimization against volumetric intensity fields and requires explicit ray-based accumulation.
%
%
A growing body of work has sought to improve 3DGS efficiency by controlling primitive density and overlap.
Prior analysis shows that densification often produces redundant primitives through cloning rather than informative splitting~\cite{huang2025decomposing}, motivating pruning-based methods~\cite{fan2024lightgaussian,hanson2024pup,zhang2024lp}. 
In particular, LightGaussian~\cite{fan2024lightgaussian}, PUP 3D-GS~\cite{hanson2024pup}, and LP-3DGS~\cite{zhang2024lp} estimate primitive importance from image-space rendering statistics to remove low-impact primitives.
%
%
%
%
However, these approaches are inherently view-dependent and are therefore not directly suited to volumetric data.
In contrast, our method formulates primitive importance directly in volumetric space by incorporating both signal contribution and structural characteristics.
Our structure-aware pruning strategy prioritizes anisotropic, high-signal primitives,
yielding compact, structurally aligned representations for sparse volumetric data, while preserving structural fidelity.

\vspace{-0.095in}
\section{Preliminaries: 3DGS}
\label{sec:3dgs}
In 3DGS, each primitive is a 3D Gaussian defined by its center, covariance, opacity, and color.
The Gaussian function is expressed as:
\vspace{-0.4em} 
\begin{equation}
G(x) = o \cdot e^{-\frac{1}{2} (x - \mu)^T \Sigma^{-1} (x - \mu)}, \quad \Sigma = RSS^T R^T,
\label{eq:gaussian}
\end{equation}
where $x \in \mathbb{R}^3$ denotes an arbitrary 3D position, $\mu \in \mathbb{R}^3$ is the Gaussian center,
$o$ is the opacity, and $\Sigma \in \mathbb{R}^{3 \times 3}$ is the covariance matrix parameterized by rotation $R$ and scale $S$.
%
%
%
%
%
%
During rendering, Gaussians are projected into screen space, and the color \(C\) at each pixel is computed via front-to-back alpha compositing:
\vspace{-0.5em} 
\begin{equation}
C = \sum_{i \in N} c_i \alpha_i \prod_{j=1}^{i-1} (1 - \alpha_j),
\label{eq:alpha_blending}
\end{equation}
%
where \(N\) is the set of Gaussians contributing to the pixel, and \(c_i\) and \(\alpha_i\) are the color and opacity of the \(i\)-th Gaussian after projection.
%
3DGS further accelerates rendering via a tile-based 2D rasterizer that depth-sorts Gaussians per tile and alpha-composites them at each pixel using shared memory parallelism.
While efficient for image-based rendering, this strategy does not model volumetric ray integration\sm{~\cite{talegaonkar2025volumetrically}}.
%
%
%
3DGS also employs ADC, increasing primitive density by cloning or splitting Gaussians based on positional gradients.
While this rasterization strategy is highly efficient for image-based rendering, it does not explicitly model volumetric signal accumulation along rays.
%
%

%
A key limitation of 3DGS arises from the spatial support of each Gaussian.
In practice, 3DGS adopts the rule of considering influence up to three standard deviations from the mean, defined as:
\vspace{-0.5em} 
\begin{equation}
    d^\top \Sigma^{-1} d \leq 9, \quad d = x - \mu,
    \label{eq:long_tail}
\end{equation}
where $d$ is the displacement vector from center $\mu$. 
%
%
%
This $3\sigma$ support often overestimates the primitive footprint, leading to excessive overlap across primitives and increased memory and rendering cost. 
As a result, 3DGS produces redundant primitives misaligned with the underlying volumetric structure, motivating alternative primitive representations and learning strategies better suited for sparse volumetric signals.
\vspace{-0.075in}
\section{Method}
Motivated by the limitations of 3DGS discussed in Section~\ref{sec:3dgs}, particularly excessive primitive overlap and limited ability to represent volumetric structures, we propose ESVR, an ellipsoid-based sparse volume rendering framework that directly learns and renders volumetric data in 3D space. 
Figure~\ref{fig:pipeline} illustrates the overall pipeline of ESVR. 
Our framework consists of two main components: (1) a structure-aware primitive learning stage that fits differentiable ellipsoids to the volumetric intensity field, and (2) a rendering stage based on per-primitive ray sampling for accurate TF evaluation.
To support large-scale volumes, we further incorporate a chunk-based training scheme with ghost ellipsoids, providing boundary context across neighboring chunks during optimization and enabling scalable learning on volumes that exceed GPU memory.
\vspace{-0.055in}
\subsection{Differentiable Ellipsoid Primitive}
\label{sec:ellipsoid}
As mentioned in Section~\ref{sec:3dgs}, long-tailed Gaussian primitives introduce excessive spatial overlap, increasing memory usage and computational cost. 
To address these limitations, we adopt constant-density ellipsoidal primitives inspired by EVER~\cite{mai2024ever}.
EVER represents primitives as indicator functions restricted to ellipsoidal boundaries:
\vspace{-0.5em} 
\begin{equation}
    P_{\text{EVER}}(x) = o \cdot \mathbf{1}_{d^\top \Sigma^{-1} d \leq 1},
    \label{eq:EVER}
\end{equation}
where $d = x - \mu$ is the displacement from the center $\mu$, and $\mathbf{1}_{\cdot}$ is the indicator function.
Compared to the conventional Gaussian support of $3\sigma$ (Eq.~\ref{eq:long_tail}), this formulation reduces the spatial footprint by up to $9\times$.
%
%
%
%
However, the hard boundary is non-differentiable, producing vanishing gradients during optimization.
EVER addresses this via ray-based accumulation, where the hard boundary is implicitly softened via transmittance integration along the ray.
However, this approach is designed for image-based scene reconstruction and is not directly applicable to optimization against a 3D intensity field. 
To enable stable training in volumetric space, we replace the hard boundary with a differentiable sigmoid formulation ensuring that the primitive itself is fully differentiable:
\vspace{-0.5em} 
\begin{equation}
    P(x) = I \cdot \sigma\big(k \cdot (1 - d^\top \Sigma^{-1} d)\big),
    \label{eq:ESVR_primitive}
\end{equation}
where $I \in \mathbb{R}$ denotes center intensity, $\sigma(\cdot)$ is the sigmoid function and $k$ controls boundary sharpness.
However, $P(x)$ exhibits reduced peak intensity when $k$ becomes small, deviating from the desired unit response at the center (Appendix Fig.~\ref{fig:1d_primitives} (a)).
To decouple peak intensity from $k$, we normalize the primitive as $\tilde{P}(x) = {P(x)} / {P(\mu)}$, preserving consistent peak opacity while allowing $k$ to control the support width.
The spatial cutoff is adjusted accordingly (Appendix Fig.~\ref{fig:1d_primitives} (b)) to avoid truncation artifacts during rendering and voxelization.
In practice, we use $k$ as a fixed hyperparameter for efficiency. Although making $k$ learnable per primitive slightly improves reconstruction quality, it introduces additional computational overhead and reduces rendering speed.
Further analysis of $k$ and its variants is provided in Appendix~\ref{app:learnable-k} and ~\ref{app:boundary-sharpness}.

\vspace{-0.095in}
\subsection{Structure-Aware Primitive Learning}
\label{sec:primitive_learning}

Structure-aware primitive learning consists of two components: 3D intensity field fitting (Sec.~\ref{sec:optimize}) and structure-aware pruning (Sec.~\ref{sec:pruning}). 
During 3D intensity field fitting, we optimize ellipsoidal primitive parameters using a reconstruction loss at every iteration, while applying densification to dynamically place and refine primitives through gradient-based splitting and cloning. 
We refer to the combination of densification and structure-aware pruning as ADC. ADC is applied during the early optimization to increase primitive density and improve signal coverage, while pruning is performed at selected iterations to remove redundant primitives. After densification, pruning is progressively applied during subsequent optimization, enabling a compact representation while preserving structural fidelity.
\vspace{-0.075in}
\subsubsection{3D Intensity Field Fitting}
\label{sec:optimize}
Unlike 3DGS, which learns view-dependent color by storing spherical harmonics (SH) coefficients and an opacity for each primitive, our framework directly models the volumetric intensity field. \sm{Here, intensity denotes the per-voxel scalar value of the original volume.}
Each primitive stores its center intensity $I$, together with geometric parameters $(\mu, \Sigma)$ as defined in Section~\ref{sec:ellipsoid}.
The predicted voxel intensity at $x$ is the sum over all primitives:
\vspace{-0.5em} 
\begin{equation}
    \hat{V}(x)\;=\;\sum_{i\in{N}} \tilde{P_i}(x)
    \;=\;\sum_{i\in{N}} I_i \frac{\,\sigma\!\Big(k\,\big(1 - d_i^\top \Sigma_i^{-1} d_i\big)\Big)}{\sigma(k)},
    \label{eq:sum_of_primitives}
\end{equation}
\sm{where $N$ is the set of primitives contributing at $x$, and $P_i(x) = \sigma\!\left(k\left(1 - d_i^{\top}\Sigma_i^{-1} d_i\right)\right)$ denotes the unnormalized profile of primitive~$i$. We normalize it by its value at the primitive center, $P_i(\mu_i)=\sigma(k)$, where $d_i=0$. Since this value is identical for every primitive, the normalization factor is simply written as the constant $\sigma(k)$.
}
%
As sampling the whole volume at once is infeasible, a subvolume $V_{\mathrm{gt}}\!\in\!\mathbb{R}^{s\times s\times s}$ is sequentially subsampled from the ground truth (GT) volume at each iteration, as illustrated in Figure~\ref{fig:pipeline} (a).
Then, we align primitive coordinates to the GT volume coordinates and voxelize only the overlapping region to obtain the prediction subvolume $\hat{V}$.
We use an L1+SSIM hybrid loss: 
\begin{equation}
    \mathit{L}_\text{rec}
    = \beta \mathit{L}_{1} + (1 - \beta)\mathit{L}_{\text{D-SSIM}}, \quad \beta\in[0,1].
    \label{eq:recon_loss}
\end{equation}

\subsubsection{Structure-aware 
Pruning}
\label{sec:pruning}


While ADC improves coverage and reconstruction accuracy, it rapidly increases the number of primitives through excessive overlap, with many contributing little information and structures that could be captured by a single elongated primitive are instead fragmented into multiple isotropic ones~\cite{huang2025decomposing}. Furthermore, elongated primitives have been shown to better preserve high-frequency structural details~\cite{kerbl20233d}.
%
%
%
%
%
%
%
%
%
As illustrated in Figure~\ref{fig:pipeline} (b), we address this by computing a structure-aware importance score for each ellipsoid at selected pruning iteration.
Unlike LightGaussian~\cite{fan2024lightgaussian}, which scores primitives from accumulated 2D ray contributions and is inherently view-dependent, our setting operates in volumetric space and requires an importance measure reflecting both signal contribution and structural relevance.
To this end, we define a structure-aware importance score that integrates both intensity and geometric properties of each primitive, given by
%
\sm{
\begin{equation}
    \mathrm{score}_i 
    = |\Omega_i| \cdot I_i \cdot \mathrm{FA}_i, \hspace{2pt}
    \Omega_i = \{ v \mid v \text{ is covered by Ellipsoid } i \},
    \label{eq:score}
\end{equation}}
where $i$ is the primitive index, 
\sm{$I_i$ is the learned center intensity of the $i$-th ellipsoid, and $|\Omega_i|$ denotes its voxel coverage. A voxel 
$v$ is considered \emph{covered} if the ellipsoid yields a non-negligible 
contribution at its center, i.e., $\tilde{P}_i(v) \geq \varepsilon$ with 
$\varepsilon = 1/255$. }
$\mathrm{FA}_i \in [0,1]$ is the fractional anisotropy 
computed from the eigenvalues $(\lambda_{1,i}, \lambda_{2,i}, \lambda_{3,i})$ of covariance matrix $\Sigma_i$.
%
Consequently, larger ellipsoids yield higher scores due to their broader support.
The FA value ranges from 0 to 1, with higher values indicating stronger anisotropy, and is calculated as
\begin{equation}
    \mathrm{FA}_i =
    \sqrt{\tfrac{1}{2}}\,
    \frac{\sqrt{(\lambda_{1,i}-\lambda_{2,i})^2 + (\lambda_{2,i}-\lambda_{3,i})^2 + (\lambda_{3,i}-\lambda_{1,i})^2}}
         {\sqrt{\lambda_{1,i}^2 + \lambda_{2,i}^2 + \lambda_{3,i}^2}}.
    \label{eq:fa}
\end{equation}
%
%
\sm{The coverage term $|\Omega_i|$ scales with the spatial footprint of a primitive,}
while FA emphasizes anisotropic primitives that align with structural boundaries and high-frequency features~\cite{kerbl20233d}. 
\sm{Moreover, since $I_i$ is optimized to reconstruct the ground-truth field through the sum of overlapping primitive contributions, redundant fragments converge to similar local intensities and individually receive low importance scores, while the single well-fitted primitive retains a high score.}
\sm{At each pruning iteration, primitives are ranked by $\mathrm{score}_i$ and the lowest-scoring ones are removed according to a preset pruning ratio.}
This effect is further analyzed in Section~\ref{sec:ablation-pruning}.
This formulation prioritizes high-intensity, elongated primitives while suppressing isotropic or redundant \sm{fragmented} ones, improving efficiency while preserving structural fidelity.

\vspace{-0.5em}
\subsection{Ghost Ellipsoids for Scalable Chunk-based Training}
\label{sec:chunk}

\begin{figure}[t]
 \centering
 \includegraphics[width=0.9\linewidth]{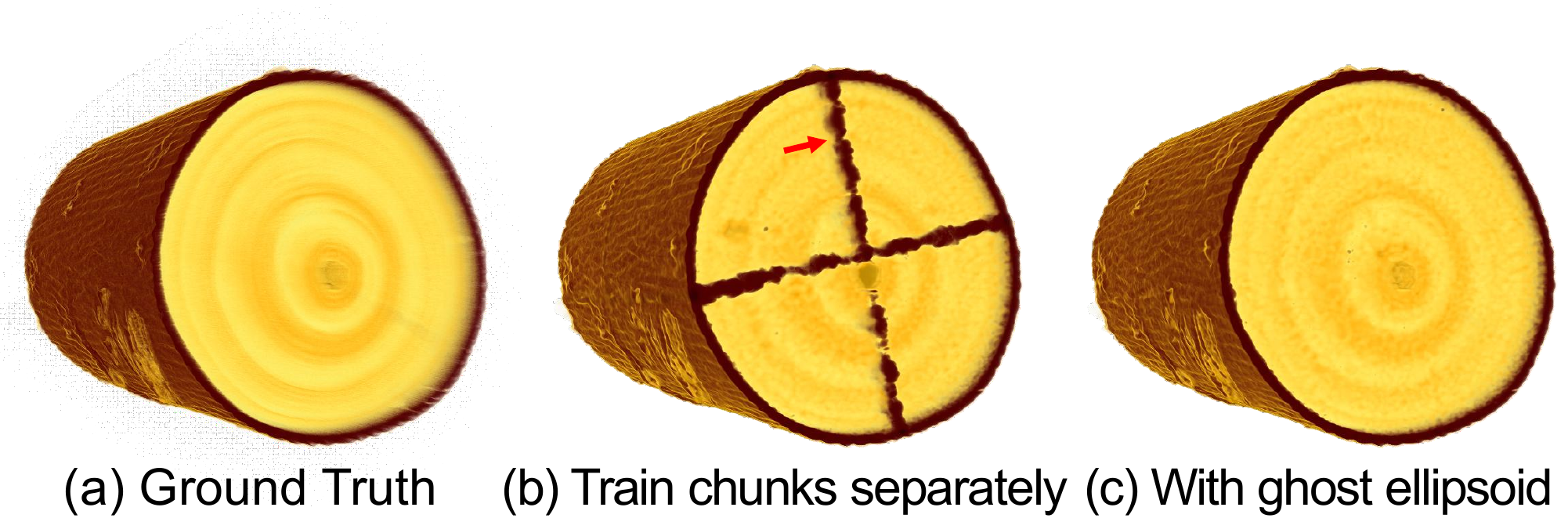}
 \vspace{-0.08in}
   \caption{
   Visualization of Woodbranch dataset: (a) ground truth, (b) training chunks separately and (c) with ghost ellipsoids in large volume learning. 
   }
 \label{fig:ghostcell}
\end{figure}
Direct access to the raw volume during training introduces two challenges for extremely large datasets: (1) volumes exceeding GPU memory cannot fit in a single-GPU setup; and (2) training time grows with volume resolution, as the voxelized subvolume used for loss computation becomes increasingly large. We address both through a divide-and-conquer strategy, partitioning the volume into spatially non-overlapping chunks and training each chunk on a dedicated GPU.
%
However, training chunks separately can 
introduce boundary artifacts, as ellipsoids near chunk borders lack support from neighboring regions (Fig.~\ref{fig:ghostcell}). Unlike chunk-based 3DGS methods~\cite{chen2024dogs,kerbl2024hierarchical,liu2024citygaussian,lin2024vastgaussian}, which resolve boundary inconsistencies by merging chunks into a global coarse model or through additional post-training stages, our volumetric setting requires boundary support during training itself without access to a global volume or merged model. 
%
%

To address this, we introduce \textit{ghost ellipsoids}, inspired by ghost cell techniques~\cite{tseng2003ghost}, but adapted for unstructured primitives. 
Unlike classical ghost cells, which copy boundary values from adjacent regular grid cells, we cannot rely on grid adjacency since our primitives are spatially unstructured. Instead, for each chunk, we identify boundary ellipsoids from all neighboring chunks (face-, edge-, and vertex-adjacent) by selecting those whose centers fall within a ghost region of width defined by the chunk boundary offset. These ghost ellipsoids are voxelized together with the current chunk's ellipsoids during loss computation but receive no gradient updates, providing boundary context without modifying neighboring chunks.
%
We support two training modes: \textit{sequential mode} for single-GPU training, where chunks are trained one at a time using ghost ellipsoids from completed neighbors, and \textit{parallel mode} for multi-GPU training, where all chunks are trained simultaneously with ghost ellipsoids periodically synchronized across GPUs.
%
Each chunk is trained in its own normalized coordinate space scaled to its spatial extent (e.g., $[-0.5, 0.5]^3$ for a volume split into $2\times2\times2$ chunks), with primitive positions stored after applying a per-chunk spatial offset that maps them directly into the shared world coordinate system. Training all chunks in a shared $[-1,1]^3$ space and applying post-hoc rescaling would require computing $\Sigma' = K\Sigma K^T = KRSS^TR^TK^T$ for each primitive, where $K$ is a chunk-dependent anisotropic scale matrix. Since $K$ and $R$ do not generally commute, this cannot be simplified by scaling $S$ alone, and correctly decomposing $\Sigma'$ back into rotation and scale parameters requires a QR decomposition per primitive, which is computationally impractical. By contrast, our approach avoids this entirely, and the resulting per-chunk primitive sets can be merged through simple concatenation without additional alignment.
\begin{figure}[t]
 \centering
 \includegraphics[width=0.9\linewidth]{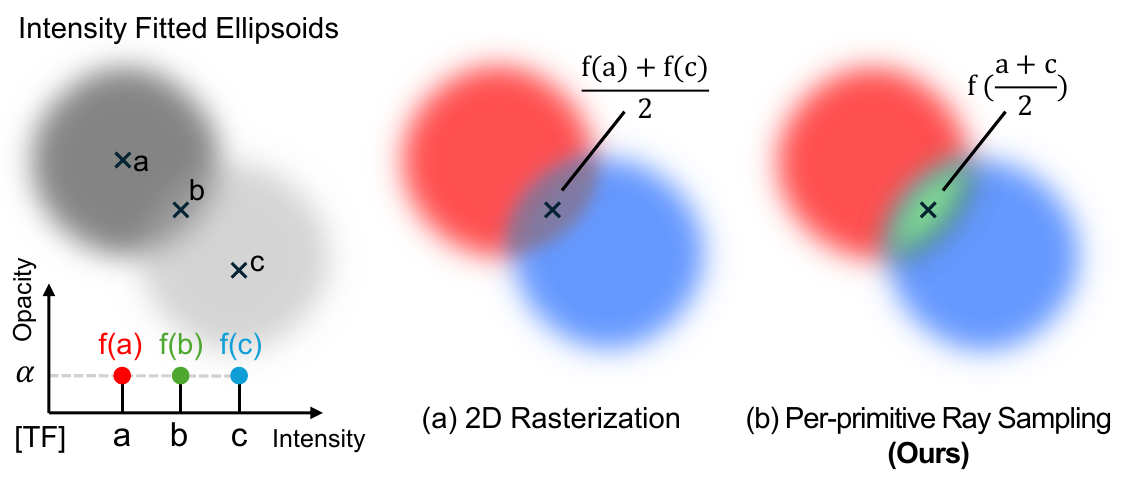}
 \vspace{-0.08in}
   \caption{
Artifacts caused by inaccurate TF mapping in 2D rasterization. Vanilla 3DGS introduces blending artifacts due to incorrect TF evaluation, while per-primitive ray sampling maintains accurate TF mapping.
   }
 \label{fig:dvrvssplat}
\end{figure}
\vspace{-0.7em}
\subsection{Per-primitive Ray Sampling}
\label{sec:ours_rendering}

\begin{figure}[t]
 \centering
 \includegraphics[width=1.0\linewidth]{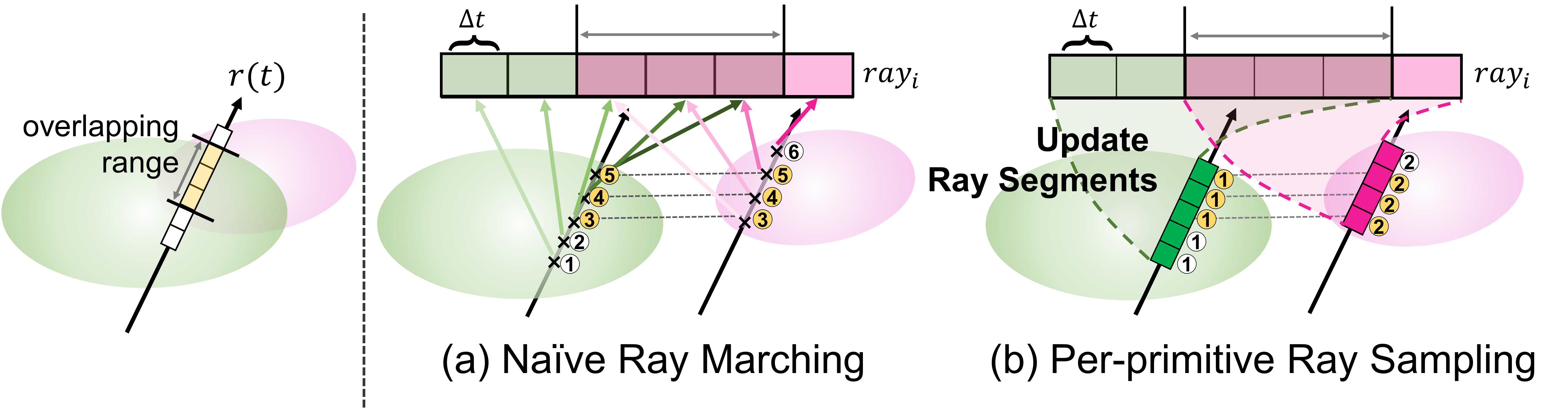}
 \vspace{-0.3in}
   \caption{ Conceptual comparison of data access patterns between (a) na\"{\i}ve ray marching and (b) per-primitive ray sampling. In ray marching, primitive contributions are queried at each ray sample, causing repeated parameter loads. In contrast, our method processes each primitive once, determines its valid depth range, and updates only the affected ray segments, reducing redundant global memory traffic.
   }
 \label{fig:ray_sampling}
\end{figure}

The standard 3DGS rendering pipeline achieves efficiency via screen-space rasterization, but this design is fundamentally incompatible with accurate TF mapping. 
As illustrated in Figure~\ref{fig:dvrvssplat} (a), the 2D pipeline blends primitive contributions in screen space after assigning colors to individual primitives (pre-classification), producing results that do not faithfully reflect the 3D intensity distribution. This limitation becomes particularly problematic in regions with overlapping primitives. 
%
Our framework instead requires precise 3D sampling of the learned primitives for accurate TF evaluation (Fig.~\ref{fig:dvrvssplat} (b)). 
We implement this via ray marching, stepping each ray through the volume and evaluating the primitive field at regular intervals.
However, naively adopting this approach incurs significant overhead, as each ray step requires checking intersections with primitives and repeatedly fetching ellipsoid parameters from memory (Fig.~\ref{fig:ray_sampling} (a)).
To address this, we propose a per-primitive ray sampling strategy where each ellipsoid independently updates only the ray segments it intersects (Fig.~\ref{fig:ray_sampling} (b)).
By organizing computation around primitives rather than rays, we avoid redundant parameter loads and accumulate contributions in 3D space, enabling both accurate TF evaluation and real-time rendering.

\noindent\textbf{Ray parameterization.} 
Primitives are first transformed by the projection matrix, and only those visible in screen space are processed.
%
%
%
For a given pixel position $(\mathbf{x}_i, \mathbf{y}_i)$, the associated ray is defined in ray space as $\mathbf{r}_i(t) = (\mathbf{x}_i, \mathbf{y}_i, 0) + t \cdot (0, 0, 1)$, where $t$ varies along the $z$-axis.
%
%
%
The ray is discretized with step size $\Delta t$ into a 1D array $ray_i[\cdot]$, where the $j$-th entry stores the accumulated intensity of the ray segment at depth $t = j \Delta t$.
Accordingly, the sample position corresponding to index $j$ is given by $\mathit{ray}_i[j] = \hat{V}(\mathbf{r}_i(j\Delta t)) = \hat{V}(\mathbf{x}_i, \mathbf{y}_i, j\Delta t)$.

\vspace{0.1em}
\noindent\textbf{Primitive contribution.}
For each primitive with center $\mu = (\mu_x, \mu_y, \mu_z)$ and covariance $\Sigma$, only the ray segments lying within its depth range are updated. 
Specifically, segments within $|t - \mu_z| \leq r_{supp} \cdot \sqrt{\Sigma'_{zz}}$ are considered, where $\sqrt{\Sigma'_{zz}}$ denotes the standard deviation of the primitive along the $z$-axis after projection, and $r_{supp}$ is the support radius factor.
%
For each valid segment, the primitive contributes as
\vspace{-0.5em} 
\begin{equation}
    \tilde{P}(\mathbf{x}_i,\mathbf{y}_i,j\cdot\Delta t) 
    = I \cdot \frac{\sigma\!\big(k \cdot (1 - \mathbf{d}^\top \Sigma^{-1} \mathbf{d})\big)}{\sigma(k)},
\end{equation}
where  $\mathbf{d} = [\mathbf{x}_i, \mathbf{y}_i, j\cdot\Delta t] - [\mu_\mathbf{x}, \mu_\mathbf{y}, \mu_\mathbf{z}]$ is the displacement from the primitive center.
%
Algorithm~\ref{alg:ray_update} summarizes the update procedure for each primitive.
\begin{algorithm}[t]
\caption{Ray Segment Update with Primitive Contribution}
\label{alg:ray_update}
\textit{$\mu, \Sigma, r$: Primitive center, covariance, and support radius} \\
\textit{$\Delta t$: sampling step} \\
\textit{${ray}_i$: ray segment } 
\begin{algorithmic}
\vspace{0.5em}
\hrule
\vspace{0.5em}
\For{each primitive}
    \State $index\_start \gets \lfloor \frac{\mu_z - r \cdot \sqrt{\Sigma_{zz}}}{\Delta t} \rfloor$
    \State $index\_end \gets \lfloor \frac{\mu_z + r \cdot \sqrt{\Sigma_{zz}}}{\Delta t} \rfloor$
    \For{$j = index\_start$ \textbf{to} $index\_end$}
        \State ${ray}_i[j] \gets {ray}_i[j] + P(\mathbf{x}_i, \mathbf{y}_i, j\cdot\Delta t)$
    \EndFor
\EndFor
\end{algorithmic}
\end{algorithm}
This eliminates explicit per-tile sorting required in 2D rasterization,  since each ray segment corresponds to a discrete depth interval and contributions accumulate in depth order.
%
Ellipsoid parameters are cached in tile-level shared memory, minimizing redundant accesses across pixels.
These optimizations accelerate rendering while maintaining accurate TF mapping.
%
%
\\
\noindent\textbf{Transfer function mapping and compositing.} 
After all primitives are processed, the accumulated ray segments $ray_i$ contain the correct 3D intensity distribution along the ray. 
These segments are traversed front-to-back, where the TF maps non-zero intensities to color and opacity, and alpha blending  (Eq.~\ref{eq:alpha_blending}) produces the final pixel color

\noindent\textbf{GPU-acceleration through intra-ray parallelism.} 
%
Although per-primitive ray sampling improves accuracy and reduces memory access and sorting overhead, updating all ray segments still incurs significant computational cost. 
To further accelerate rendering, we adopt intra-ray parallelization that divides each ray into a fixed number of depth bins.
Within each bin, primitive contributions are accumulated into intermediate color and transmittance values, then composited front-to-back yielding the same result as per-segment accumulation with significantly fewer redundant operations.

\begin{figure}[t]
 \centering
 \includegraphics[width=1.0\linewidth]{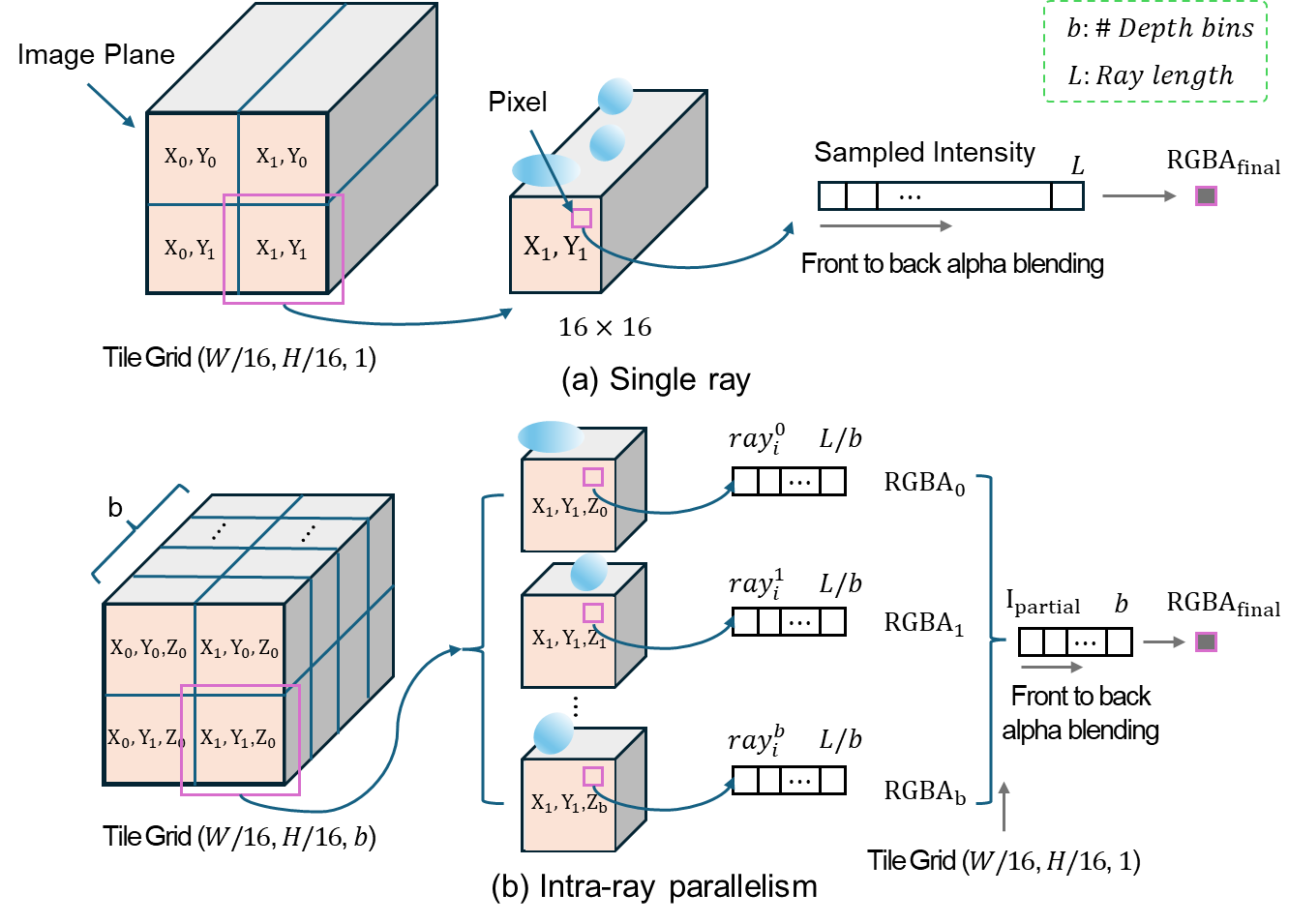} 
 \vspace{-0.3in}
   \caption{Illustration of intra-ray parallelism.}
 \label{fig:intra-ray}
\end{figure}

%
%
As illustrated in Figure ~\ref{fig:intra-ray}, we extend the conventional CUDA tile grid beyond the image plane $(x,y)$ to also span the $z$-axis, by partitioning the $z$-axis into $b$ depth bins.
This splits each pixel ray $\mathbf{r}_i$ into $b$ partial rays. 
Each primitive is mapped to the 3D tiles it intersects, instead of 2D. 
For each primitive, we enumerate intersecting 3D tiles, generating tile--primitive keys that are used to assign primitives to depth-binned tiles. For each tile, partial color and opacity values are accumulated through per-primitive ray sampling, so each pixel holds $b$ partial results. Finally, the tile grid is flattened back to $(x,y,1)$ and the partial results are composited in front-to-back order across depth bins to yield the final pixel values. See Appendix Algorithm~\ref{alg:gpu} for details.
This rendering framework extends the CUDA-based rendering pipeline of 3DGS and is integrated into our ESVR viewer for interactive TF editing and visualization (see Appendix Fig.~\ref{fig:tf_viewer}). 
\begin{table}[b]
\centering
\caption{Dataset statistics.}
\vspace{-0.1in}
\renewcommand{\arraystretch}{1.15}
\resizebox{1.0\linewidth}{!}{
\begin{tabular}{l |ccccc}
\noalign{\hrule height 1.0pt}
\textbf{Dataset} & \textbf{Resolution} & \textbf{Size} & \textbf{\makecell{Non-Zero\\ Voxel}} & \textbf{\makecell{10\% Max \\Occupancy}} & \textbf{\makecell{\sm{Screen} \\ \sm{Occupancy}}}\\
\hline
Aneurysm    & $256\times256\times256$ & 65.5MB & 0.64\%  & 0.64\%&  23\%\\
Breast      & $448\times448\times160$ & 125.4MB & 3.48\% & 3.48\%& 27\%\\
Neuron      & $2537\times1295\times133$ & 1.7GB & 98.41\% & 1.84\%& 31\%\\
BigBrain        & $1749\times1643\times1428$ & 15.1GB & 29.46\% & 29.43\%& 53\%\\
Woodbranch      & $2048\times2048\times2048$ & 32GB & 84.89\% & 3.29\%& 54\%\\
Hemibrain (vDeltaA) & $4072\times4372\times4171$ & 259GB & 0.30\% & 0.29\%& 23\%\\
Hemibrain (lLN2P) & $5115\times4774\times6128$ & 568GB  & 0.34\% & 0.34\% & 27\%\\
\hline
Beechnut   & $1024\times1024\times1546$ & 6.48GB  & 100.00\% & 100.00\%& 55\%\\
Pigheart   & $2048\times2048\times2612$ & 43.82GB & 100.00\% & 84.60\% & 68\%\\
\noalign{\hrule height 1.0pt}
\end{tabular}
}
\label{tab:dataset_info}
\end{table}

\begin{table*}[h]
\centering
\scriptsize
\renewcommand{\arraystretch}{1.0}
\setlength{\tabcolsep}{2pt}

\caption{
Quantitative comparison across seven volumetric datasets. Results are reported under two configurations: Size-M and PSNR-M—constraining storage size and reconstruction quality to match ours, respectively.
Metrics include model size (MB), 3D PSNR, and rendering speed (FPS). %
Best results are highlighted in bold. \sm{Numbers in parentheses indicate primitive counts in thousands. OOM denotes out of memory.}
}
\label{tab:volumetric_comparison}
\vspace{-0.09in}
\resizebox{0.85\textwidth}{!}{
\setlength{\tabcolsep}{3pt}
\begin{tabular}{ll | ccc | ccc | ccc | ccc | ccc | c}
\noalign{\hrule height 1.0pt}

& &
\multicolumn{3}{c|}{\textbf{fV-SRN}} &
\multicolumn{3}{c|}{\textbf{NGP}} &
\multicolumn{3}{c|}{\textbf{AMGSRN++}} &
\multicolumn{3}{c|}{\textbf{InstantVNR}} &
\multicolumn{3}{c|}{\textbf{Ours}} &
\multicolumn{1}{c}{\textbf{\sm{DVR}}} \\
\cline{3-5}\cline{6-8}\cline{9-11}\cline{12-14}\cline{15-17}\cline{18-18}
\textbf{Dataset}
& \textbf{Config} &
Size & PSNR & FPS &
Size & PSNR & FPS &
Size & PSNR & FPS &
Size & PSNR & FPS &
Size & PSNR & FPS &
FPS\\
\hline
\noalign{\vskip 0.3pt}

\multirow{2}{*}{{Aneurysm}}
& Size-M
& 2.9 & 36.89  & 0.61
& 3.0 & 43.50 & 1.12
& 2.8 & 40.08 & 0.98
& 2.8 & 39.42 & 25.08 
&\multirow{2}{*}{\textbf{2.6 (59K)}} & \multirow{2}{*}{\textbf{51.54}} & \multirow{2}{*}{\textbf{223.19}} 
& \multirow{2}{*}{137.84} \\
 & PSNR-M
& 64.0 & 42.35 & 0.66
& 62.0 & 46.17 & 1.08
& 64.0 & 43.67 & 0.53
& 66.3 & 43.29 & 16.30
& & & &\\
\hline

\multirow{2}{*}{{Breast}}
& Size-M
& 2.9 & 36.87 & 0.90
& 3.0 & 41.19 & 1.48
& 3.3 & 40.22 & 1.35
& 3.1 & 39.79 & 21.78
& \multirow{2}{*}{\textbf{2.9 (67K)}} & \multirow{2}{*}{\textbf{46.00}} & \multirow{2}{*}{\textbf{213.49}}
& \multirow{2}{*}{211.62} \\
 & PSNR-M
& 512 & 45.81 & 0.94
& 318 & 45.88 & 1.55
& 519 & 45.72 & 0.80
& 28.4 & 44.91 & 19.64
& & & &\\
\hline

\multirow{2}{*}{{Neuron}}
& Size-M
& 3.3 & 38.33 & 1.77
& 3.0 & 38.97 & 2.87
& 3.3 & \textbf{40.18} & 2.25
& 2.9 & 38.29 & 24.63
& \multirow{2}{*}{\textbf{2.8 (65K)}} & \multirow{2}{*}{39.56} & \multirow{2}{*}{\textbf{131.08}} 
& \multirow{2}{*}{90.84} \\
 & PSNR-M
& 4.0 & 39.24 & 1.82
& 4.2 & 39.21 & 2.67
& 2.9 & 39.98 & 2.25
& 15.2 & 39.28 & 18.09 
& & & &\\
\hline

\multirow{2}{*}{{BigBrain}}
& Size-M
& 7.6 & 21.62 & 0.10
& 8.2 & 22.46 & 0.18
& 8.6 & \textbf{22.75} & 0.12
& 7.3 & 20.71 & 19.61
& \multirow{2}{*}{7.6 (176K)} & \multirow{2}{*}{21.53} & \multirow{2}{*}{\textbf{48.64}} 
& \multirow{2}{*}{14.44} \\
 & PSNR-M
& 8.8 & 21.74 & 0.10
& \textbf{5.1} & 21.86 & 0.18
& 6.5 & 22.51& 0.14
& 14.3 & 21.58 & 15.05 
& & & &\\
\hline

\multirow{2}{*}{{Woodbranch}}
& Size-M
& 4.4 & 41.56 & 0.07
& 4.6 & \textbf{41.72} & 0.11
& 4.5 & 41.63 & 0.08
& 4.3 & 41.54 & 34.31
& \multirow{2}{*}{4.3 (99K)} & \multirow{2}{*}{40.60} & \multirow{2}{*}{\textbf{43.42}}
& \multirow{2}{*}{\textsc{oom}} \\
 & PSNR-M
& 3.6 & 41.55 & 0.07
& 4.1 & 41.70 & 0.12
& 3.9 & 41.59 & 0.09
& \textbf{3.5} & 41.41 & 32.00 
& & & &\\
\hline

\multirow{2}{*}{{\makecell[l]{Hemibrain\\ (vDeltaA)}}}
& Size-M
& 4.2 & 34.40 & 0.02
& 4.7 & 35.01 & 0.02
& 4.4 & 34.69 & 0.03
& - & - & -
&\multirow{2}{*}{\textbf{4.2 (99K)}} & \multirow{2}{*}{\textbf{35.34}} & \multirow{2}{*}{\textbf{137.95}}
& \multirow{2}{*}{\textsc{oom}} \\
& PSNR-M
& 17.9 & 34.97 & 0.02
& 5.5 & 35.29& 0.02
& 4.93 & 35.30 & 0.02
& - & - & - 
& & & &\\
\hline

\multirow{2}{*}{{\makecell[l]{Hemibrain\\ (lLN2P)}}}
& Size-M
& 6.1 & 35.12 & 0.02
& 6.2 & 36.56 & 0.02
& 6.1 & 36.55 & 0.02
& 7.7 & 30.24 & 8.74
&\multirow{2}{*}{\textbf{ 5.8 (136K)}} & \multirow{2}{*}{\textbf{37.03}} &\multirow{2}{*}{\textbf{86.76}} 
& \multirow{2}{*}{\textsc{oom}} \\
& PSNR-M
& 16.0 & 36.82 & 0.02
& 5.4 & 36.67& 0.02
& 6.1 & 36.55 & 0.02
& 7.7 & 30.24 & 8.74
& & & &\\

\noalign{\hrule height 1.0pt}
\end{tabular}}
\end{table*}


\begin{table*}[t]
\centering
\caption{2D rendering results under the SIZE constraint. Higher PSNR/SSIM and lower LPIPS indicate better quality. The best and second-best results are highlighted in bold and underline, respectively. }
\label{tab:2dpsnr}
\vspace{-0.1in}
\resizebox{\textwidth}{!}{
\setlength{\tabcolsep}{2.5pt}
\scriptsize
\begin{tabular}{l |ccc | ccc | ccc | ccc | ccc | ccc}
\noalign{\hrule height 1.0pt}
& \multicolumn{3}{c|}{\textbf{fV-SRN}} 
& \multicolumn{3}{c|}{\textbf{NGP}}
& \multicolumn{3}{c|}{\textbf{AMGSRN++}}
& \multicolumn{3}{c|}{\textbf{InstantVNR}}
& \multicolumn{3}{c|}{\textbf{Ours (w/o pruning)}}
& \multicolumn{3}{c}{\textbf{Ours}} \\
\cline{2-4} \cline{5-7} \cline{8-10} \cline{11-13} \cline{14-16} \cline{17-19}
\textbf{Dataset}
& PSNR & SSIM & LPIPS
& PSNR & SSIM & LPIPS
& PSNR & SSIM & LPIPS
& PSNR & SSIM & LPIPS
& PSNR & SSIM & LPIPS
& PSNR & SSIM & LPIPS \\
\hline
Aneurysm
& 23.67 & 0.8900 & 0.1098
& 27.86 & 0.9416 & 0.0676
& 25.90 & 0.9177 & 0.0946
& 22.47 & 0.8953 & 0.1088
& \textbf{37.45} & \textbf{0.9905} & \textbf{0.0102}
& \underline{31.58} & \underline{0.9676} & \underline{0.0402} \\

Breast
& 24.38 & 0.8635 & 0.0974
& 26.26 & 0.8904 & 0.0803
& 25.15 & 0.8834 & 0.0864
& 27.62 & 0.8921 & 0.0831
& \textbf{33.05} & \textbf{0.9590} & \textbf{0.0413}
& \underline{32.22} & \underline{0.9493} & \underline{0.0504} \\

Neuron
& 15.95 & 0.2805 & 0.6001
& 16.60 & 0.3485 & 0.5399
& 18.59 & 0.5113 & 0.4518
& 15.50 & 0.2982 & 0.5967
& \textbf{19.32} & \textbf{0.5827} & \textbf{0.3538}
& \underline{18.62} & \underline{0.5239} & \underline{0.4302} \\

BigBrain
& 21.89 & 0.6560 & 0.2644
& 22.40 & 0.6682 & \textbf{0.2374}
& \textbf{22.69} & \textbf{0.6748} & \underline{0.2383}
& 22.19 & 0.6589 & \underline{0.2404}
& 19.91 & 0.6328 & 0.2469
& \underline{20.68} & \underline{0.6281} & 0.2412 \\

Woodbranch
& 31.02 & 0.8795 & 0.1913
& \underline{32.00} & \textbf{0.9148} & \textbf{0.1276}
& 31.38 & 0.8910 & 0.1969
& 31.97 & \underline{0.8954} & \underline{0.1490}
& 29.04 & 0.8458 & 0.2194
& \textbf{32.50} & 0.8782 & 0.1812 \\

Hemibrain (vDeltaA)
& 22.01 & 0.8505 & 0.1455
& 20.42 & 0.7873 & 0.2426
& 21.87 & 0.8520 & \textbf{0.1334}
& - & - & -
& \textbf{23.93} & 0.8329 & 0.2745
& \underline{23.03} & \textbf{0.8630} & \underline{0.1730} \\

Hemibrain (lLN2P)
& 20.06 & 0.8157 & 0.1833
& 22.35 & 0.8755 & 0.1326
& 22.16 & 0.8737 & 0.1354
& 15.24 & 0.5819 & 0.4001
& 22.75 & 0.8904 & 0.1034
& \textbf{23.54} & \textbf{0.9023} & \textbf{0.0953} \\
\noalign{\hrule height 1.0pt}
\end{tabular}
}
\end{table*}

\vspace{-0.5em}
\section{Experiments}
\subsection{Setup}
\noindent\textbf{Dataset.} 
We evaluate on the datasets in Table~\ref{tab:dataset_info}, covering raw single-precision floating-point volumes ranging from 65.5MB to 568GB. Datasets are categorized as small($<1GB$), medium($\sim1GB$), large($>10GB$), and extremely large($>100GB$). 
Sparsity is assessed by two metrics: non-zero voxel ratio and occupancy of voxels exceeding 10\% of the maximum intensity. 
Aneurysm, Breast, and Hemibrain\cite{scheffer2020connectome} are sparse by both metrics. Neuron, BigBrain~\cite{amunts2013bigbrain}, and Woodbranch have high non-zero voxel ratios but low high-intensity occupancy, indicating they are dense in occupancy but sparse in meaningful signal. \sm{We also report screen occupancy, the average fraction of rendered pixels covered by the volume over a fixed 360° camera trajectory.}
Beechnut and Pigheart are truly dense; while our primary focus is on sparse volumes, we include dense datasets for completeness.
%
The Neuron dataset is a private dataset provided by our collaborators. 
See Appendix~\ref{app:dataset-details} for dataset details.
\noindent\textbf{Implementation.}
Our system is built upon the 3DGS framework~\cite{kerbl20233d} and the CUDA voxelizer from $\text{R}^2-$Gaussian~\cite{zha2024r}.
All training is conducted on NVIDIA RTX A6000 GPUs, and evaluation on an \sm{NVIDIA} RTX 3090. 
For volumes $>10$GB, we apply chunk-based training (8–125 chunks depending on dataset). 
We set $k=5$, cutoff to $1.42$, and $\beta=0.2$. 
Each chunk (or full volume) is initialized with 100,000 points \sm{and partitioned into cubic subvolumes whose side length is approximately half its longest axis; subvolumes are traversed in a fixed sequential order} and one epoch corresponds to a full traversal of subvolumes. Therefore, epoch counts vary with volume size.
Details are provided in Appendix~\ref{app:imp}.

\noindent\textbf{Rendering and evaluation.}
After training, volumes are rendered using per-primitive ray sampling at $800\times800$ resolution with 1 voxel sampling step.
FPS is measured over a full 360\degree rotation (0.5\degree increments, 720 views).
We report PSNR and SSIM (higher is better), and LPIPS~\cite{zhang2018perceptual} (lower is better) for perceptual similarity. Rendering speed is measured using the full rendered view. Therefore, the multi-scale rendering framework used for acceleration is not applied to any of the comparison methods.

\begin{figure*}[t]
 \centering
 \includegraphics[width=1.0\linewidth]{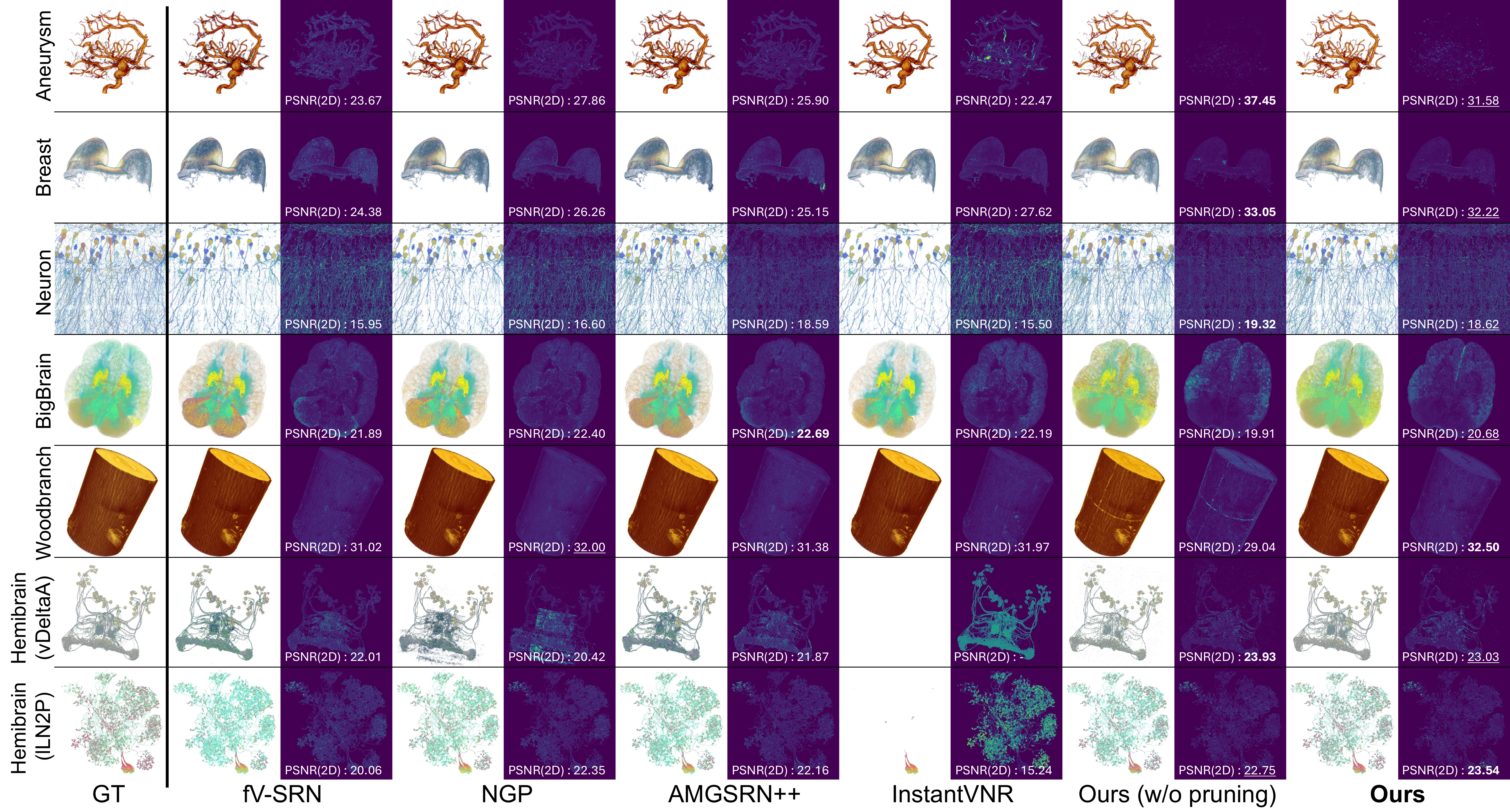}
    \vspace{-0.15in}
   \caption{Qualitative comparison of TF-mapped renderings and \sm{absolute per-pixel} error maps with respect to DVR ground truth under a model-size–matched setting. Woodbranch and Hemibrain are downsampled ($\times 8$ and $\times 64$) to enable DVR (GT) rendering without OOM.
   }
 \label{fig:rendered_vol}
\end{figure*}


\noindent\textbf{Baseline comparison.}
We compare our method against four SRN-based methods: fV-SRN, NGP, AMGSRN++, and InstantVNR. \sm{While NGP and InstantVNR share a similar backbone, InstantVNR incorporates out-of-core sampling for training large volumetric datasets, motivating its inclusion as a separate baseline.} 
Each baseline is trained following its original protocol, with model size or PSNR matched for fair comparison.
We use the neural volume renderer from AMGSRN++ to evaluate the rendering performance of fV-SRN, NGP, and AMGSRN++. For InstantVNR, we use its own framework with ray marching, sample streaming, and macro-cell acceleration.
For volumes $>$10GB, fV-SRN and NGP exceed single-GPU memory limits; we therefore use the domain decomposition implementation from AMGSRN++ for both methods, and InstantVNR’s out-of-core sampling, enabling comparison on the largest datasets. Feature grid compression is not used for AMGSRN++.
We additionally compare rendering performance against standard DVR and RTX-accelerated DVR with empty space skipping (denoted as DVR (RTX))~\cite{wald2021faster}, and further evaluate our method in a resource-constrained environment 
using a laptop GPU (NVIDIA RTX 4050 Laptop GPU).
For DVR, we use the GPU-based volume rendering mode in ParaView~\cite{ahrens2005paraview}.
Image-based 3DGS methods are discussed qualitatively in Section~\ref{sec:discussion}, as they do not support 3D PSNR evaluation or flexible TF mapping.

\vspace{-0.5em}
\subsection{Quantitative Comparison}%

\noindent\textbf{Compression and reconstruction quality.} 
Table~\ref{tab:volumetric_comparison} presents a quantitative comparison across all datasets. We report results under two configurations: Size-M (model size matched to our method) and PSNR-M (reconstruction quality matched to our method). Compression performance varies with dataset sparsity. 
For high sparsity datasets (Aneurysm, Breast, Hemibrain), our method consistently outperforms all baselines under Size-M setting, achieving up to 8–12 dB higher 3D PSNR.
Under PSNR-M setting, competing methods require substantially larger models, often 10 to 20$\times$ larger, to reach comparable quality. In some cases, they still fail to match our results.
For datasets with lower sparsity (e.g., BigBrain and Woodbranch), our method achieves comparable reconstruction quality to SRN-based methods such as NGP and AMGSRN++, which are better suited for modeling smooth and dense signals. 
For large-scale Hemibrain datasets exceeding single-GPU memory, InstantVNR relies on out-of-core training. However, as noted by the authors, its random sampling strategy is ill-suited for sparse volumes and fails to capture thin structures. This is particularly evident in Hemibrain (vDeltaA), which contains finer structures than (lLN2P). InstantVNR fails to achieve meaningful reconstruction quality in the former and limited performance in the latter, while our method maintains consistent reconstruction quality across both.
Overall, our primitive-based representation is particularly effective for sparse volumetric data, while remaining competitive on denser datasets.
%
%
%
%
\begin{table}[t]
\centering
\setlength{\tabcolsep}{5pt}
\renewcommand{\arraystretch}{1.0}
\caption{Rendering performance and VRAM usage comparison across datasets. \sm{Measured on RTX~3090, except \textit{Ours~(Laptop)} (RTX~4050).}}
\vspace{-0.5em}
\resizebox{\linewidth}{!}{
\begin{tabular}{ll|cccc}
\noalign{\hrule height 1.0pt}
\textbf{Dataset} & \textbf{Metric} & \textbf{Ours} & \textbf{Ours (Laptop)} & \textbf{DVR} & \textbf{DVR (RTX)} \\
\hline
\multirow{2}{*}{Aneurysm}
& FPS  & 223.19 & 125.08 & 137.84 & \textbf{225.27} \\
& VRAM & 1.64GB & 1.43GB & 1.49GB & 1.77GB \\
\hline
\multirow{2}{*}{Breast}
& FPS  & 213.49 & 118.25 & 211.62  & \textbf{347.20} \\
& VRAM & 1.64GB & 1.43GB & 1.51GB   & 1.79GB \\
\hline
\multirow{2}{*}{Neuron}
& FPS  & 131.08 & 90.73 & 90.84 & \textbf{164.69} \\
& VRAM & 1.64GB & 1.42GB & 2.26GB & 2.67GB \\
\hline
\multirow{2}{*}{BigBrain}
& FPS  & \textbf{48.64}  & 24.60 & 14.44 & 40.27 \\
& VRAM & 1.74GB & 1.54GB & 5.43GB & 5.83GB \\
\hline
\multirow{2}{*}{Woodbranch}
& FPS  & \textbf{43.42}  & 19.51 & \textsc{oom}      & 33.91 \\
& VRAM & 1.71GB & 1.50GB & \textsc{oom}      & 18.2GB \\
\hline

\multirow{2}{*}{\begin{tabular}[c]{@{}l@{}}Hemibrain\\(vDeltaA)\end{tabular}}
& FPS  & \textbf{137.95} & 100.11 & \textsc{oom}      & \textsc{oom} \\
& VRAM & 1.69GB & 1.44GB & \textsc{oom}      & \textsc{oom} \\
\hline
\multirow{2}{*}{\begin{tabular}[c]{@{}l@{}}Hemibrain\\(lLN2P)\end{tabular}}
& FPS  & \textbf{86.76}  & 77.64 & \textsc{oom}      & \textsc{oom} \\
& VRAM & 1.67GB & 1.46GB & \textsc{oom}      & \textsc{oom} \\
\noalign{\hrule height 1.0pt}
\end{tabular}
}
\label{tab:rendering_performance}
\end{table}









\noindent\textbf{Rendering performance.} Our method consistently achieves real time rendering across all datasets, while InstantVNR, despite being a dedicated acceleration of Instant-NGP for volume rendering, frequently falls below interactive frame rates on large volumes. On small to medium datasets, our renderer achieves 223, 213, and 131 FPS, well above interactive rates.
On larger and relatively dense datasets (BigBrain, Woodbranch), our method outperforms InstantVNR while maintaining interactive rendering performance. 
As volume resolution increases to hundreds of gigabytes (e.g., Hemibrain), InstantVNR drops to low frame rates (e.g., ~8 FPS), whereas our method sustains interactive performance (achieving 137 FPS and 86 FPS on Hemibrain (vDeltaA) and (lLN2P), respectively).
As shown in Table~\ref{tab:2dpsnr}, InstantVNR exhibits lower SSIM and higher LPIPS, especially on large-scale sparse datasets such as Hemibrain. In contrast, our method achieves competitive or superior 2D quality across all metrics without sacrificing speed.

Table~\ref{tab:rendering_performance} further compares rendering performance across a laptop GPU, standard DVR, and DVR (RTX) with empty space skipping. Our method maintains high frame rates on a mobile GPU with significantly less VRAM. Compared to conventional DVR, we consistently achieve higher frame rates across all datasets. Although DVR (RTX) outperforms our method on small sparse volumes, where hardware ray traversal is most effective, its memory usage scales poorly with resolution. DVR runs out of memory starting from Woodbranch, and DVR (RTX) from Hemibrain onward. In contrast, our method maintains a stable and low VRAM footprint across all tested datasets.
%
%
%
Overall, these results demonstrate that our primitive-based framework compresses volumes of up to hundreds of gigabytes into compact representations while retaining competitive or superior PSNR on large sparse data, and simultaneously achieves rendering speeds well beyond real time across all tested datasets. Together, the compression efficiency and rendering performance make our framework well-suited for interactive visualization of large-scale volumetric data. 

\sm{
Additional experiments are provided in the Appendix, including comparisons with image-based baselines (LightGaussian, $R^2$-Gaussian), traditional compression methods and VDB-based methods.}

\vspace{-0.3em}
\subsection{Qualitative Comparison}
Figure~\ref{fig:rendered_vol} presents TF-mapped renderings and corresponding error maps under the Size-M setting. 
%
For sparse vascular structures (Aneurysm, Breast), our method preserves fine tubular structures that SRN-based methods tend to over-smooth.
%
For datasets with high-frequency details and sparse occupancy (Neuron), our method achieves lower overall error, better preserving sharp intensity variations and thin structures. 
For BigBrain, which has the highest 10\% max occupancy among our datasets, SRN-based methods achieve better overall reconstruction quality. 
However, our method shows lower error in thin and complex regions. 
For Woodbranch, despite its dense occupancy, our method remains competitive and often produces better renderings around surfaces and edges, where our primitive-based representation is advantageous. 
For the Hemibrain datasets, INR-based methods produce degraded results, failing to recover sparse structures. 
In particular, InstantVNR partially reconstructs Hemibrain (lLN2P) but fails on finer structures in Hemibrain (vDeltaA),
while
our method preserves fine structural details across both.
These observations are mostly consistent with the quantitative results in Table~\ref{tab:volumetric_comparison} and highlight the complementary strengths of the two approaches: our primitive-based representation excels on sparse, high-frequency volumes, while continuous neural representations remain competitive on dense and smooth data.
\begin{table}[h]
\centering
\caption{\sm{Training time comparison across datasets in Size-M setting. 
For large datasets, we report per-chunk average training time with the number of chunks in parentheses. Full training time is reported for InstantVNR, which uses out-of-core training.}
}

\footnotesize

\setlength{\tabcolsep}{4pt}
\renewcommand{\arraystretch}{1.1}

\resizebox{\columnwidth}{!}{
\begin{tabular}{l|ccccc}
\noalign{\hrule height 1.0pt}
\textbf{Dataset} & \textbf{fV-SRN} & \textbf{NGP} & \textbf{AMGSRN++} & \textbf{InstantVNR} & \textbf{Ours} \\
\hline
Aneurysm & 58s & 1m 52s & 3m 49s & 2m 18s & 6m 28s \\
Breast   & 1m 24s & 1m 22s & 3m 58s & 2m 33s & 12m 36s \\
Neuron   & 1m 21s & 1m 21s & 4m 25s & 4m 22s & 22m 0s \\

\hline
BigBrain (8) & 58s & 1m 1s  & 4m 36s & 1h 49m 54s & 1h 7m 20s \\
Woodbranch (8) & 1m 54s & 1m 36s & 4m 18s & 1h 54m 39s & 1h 34m 36s \\
Hemibrain (vDeltaA) (64) & 1m 45s & 1m 58s & 1m 59s & - & 49m 58s \\
Hemibrain (lLN2P) (125) & 1m 24s & 1m 36s & 3m 36s & 8h 29m 15s & 58m 42s \\

\noalign{\hrule height 1.0pt}
\end{tabular}
}
\label{tab:training_time}
\end{table}

\begin{table}[h]
\centering
\scriptsize
\renewcommand{\arraystretch}{1.0}
\setlength{\tabcolsep}{3pt}

\caption{
Quantitative comparison on two dense volumetric datasets under Size-M setting.
}
\label{tab:dense}
\resizebox{1.0\linewidth}{!}{

\begin{tabular}{l | ccc | ccc | ccc | ccc}
\noalign{\hrule height 1.0pt}

& \multicolumn{3}{c|}{\textbf{NGP}} 
& \multicolumn{3}{c|}{\textbf{AMGSRN++}} 
& \multicolumn{3}{c|}{\textbf{InstantVNR}} 
& \multicolumn{3}{c}{\textbf{Ours}} \\
\cline{2-4}\cline{5-7}\cline{8-10}\cline{11-13} 

\textbf{Dataset}
& Size & PSNR & FPS
& Size & PSNR & FPS
& Size & PSNR & FPS
& Size & PSNR & FPS \\
\hline
\noalign{\vskip 0.3pt}

Beechnut
& 1.9 & \textbf{45.46} & 0.50
& 1.8 & 44.97 & 0.46
& 3.4 & 44.54 & 26.39
& 1.6 & 39.62 & \textbf{43.08}\\


Pigheart
& 14.4 & 38.08 & 0.15
& 15.2 & \textbf{38.11} & 0.13
& 14.3 & 37.57 & 18.22
& 14.3 & 31.95 & \textbf{25.10}\\

\noalign{\hrule height 1.0pt}
\end{tabular}
}
\end{table}
\sm{
\subsection{Training Time}
Table~\ref{tab:training_time} summarizes the training time across datasets under the Size-M setting. 
Per-chunk average training time is reported for large datasets, except for InstantVNR where full training time is reported due to its out-of-core training strategy. 
%
On small and medium datasets (Aneurysm, Breast, Neuron), our method shows longer training times than SRN-based approaches, reflecting the cost of direct volumetric optimization. This gap becomes more pronounced for larger datasets. 
While our chunk-based optimization enables multi-GPU parallelization, in a single-GPU setting, the total training time scales with the number of chunks, leading to an increase in overall training time as volume size grows (e.g., 22m for Neuron vs. 1h 7m for BigBrain).
In contrast, InstantVNR does not adopt explicit chunking and instead processes the full volume using out-of-core sampling, resulting in substantially longer training times on large datasets (e.g., 1h49m on BigBrain and 8h29m on Hemibrain (lLN2P)) compared to other SRN-based methods. 
Overall, our method incurs higher training cost than SRN-based models but scales more effectively than out-of-core approaches for extremely large volumes under multi-GPU settings.
}

\subsection{Dense Dataset}
Although ESVR targets sparse volumetric data, we also evaluate our method on dense scientific datasets (e.g., Pigheart, Beechnut) to assess its scalability (Table ~\ref{tab:dense}). 
Unlike our main datasets, these volumes contain 100\% non-zero regions. As expected, our method achieves lower reconstruction quality compared to network-based 
approaches, which are better suited for modeling dense, continuous structures and our advantage of selectively placing primitives is less pronounced. 
Nevertheless, our method maintains near real-time rendering performance, offering a practical trade-off between reconstruction quality and efficiency.

\subsection{Ablation Study}
For a fair comparison, we adjusted the densification threshold to roughly match the number of primitives across methods. As a result, some datasets show primitive counts that differ from those reported in Table~\ref{tab:volumetric_comparison}.



\begin{table}[t!]
\centering
\small
\caption{
Ablation study on primitive types (Gaussian vs.\ Ellipsoid) before and after pruning across three datasets. Numbers in parentheses indicate primitive counts. Avg PrimHits denotes the average number of primitives per voxel, computed over voxels with non-zero primitive hits.
}
\renewcommand{\arraystretch}{0.9}
\setlength{\tabcolsep}{4pt}
\resizebox{0.9\linewidth}{!}{
\begin{tabular}{l|l|cc|cc}
\noalign{\hrule height 1.0pt}
& & \multicolumn{2}{c|}{\textbf{Ellipsoid}} & \multicolumn{2}{c}{\textbf{Gaussian}} \\
\textbf{Dataset} & \textbf{Metric} & \textit{Before} & \textit{After} & \textit{Before} & \textit{After} \\
\hline
\multirow{5}{*}{Aneurysm}
& Size (MB) & 4.03 (93k) & 1.37 (31k) & 4.11 (95k) & 1.40 (32k) \\
& 3D PSNR (dB) & 51.20 & 50.70 & 44.97 & 48.16 \\
& \sm{3D SSIM} & 0.9986 & 0.9976 & 0.9979 & 0.9977 \\
& FPS & 66.45 & 248.92 & 6.02 & 34.06 \\
& Avg PrimHits & 20.10 & 9.05 & 34.73 & 19.96 \\
\hline
\multirow{5}{*}{Breast}
& Size (MB) & 8.06 (187k) & 2.74 (63k) & 8.06 (187k) & 2.74 (63k) \\
& 3D PSNR (dB) & 48.40 & 45.62 & 46.05 & 45.04 \\
& \sm{3D SSIM}  & 0.9983 & 0.9964 & 0.9975 & 0.9965 \\
& FPS & 83.82 & 214.14 & 8.99 & 23.08 \\
& Avg PrimHits & 13.69 & 6.10 & 24.00 & 11.24 \\
\hline
\multirow{5}{*}{BigBrain}
& Size (MB) & 5.09 (118k) & 1.73 (40k) & 5.05 (117k) & 1.72 (39k) \\
& 3D PSNR (dB) & 18.79 & 19.12 & 18.00 & 18.30 \\
&\sm{3D SSIM}  & 0.8129 & 0.8147 & 0.8000 & 0.8071 \\
& FPS & 145.69 & 172.68 & 6.63 & 13.69 \\
& Avg PrimHits & 6.48 & 5.37 & 22.49 & 12.63 \\
\noalign{\hrule height 1.0pt}
\end{tabular}
}
\label{tab:ellipsoid_vs_gaussian}
\end{table}
%




\noindent\textbf{Primitive design.}
\label{sec:ablation-primitives}
Table~\ref{tab:ellipsoid_vs_gaussian} compares ellipsoid and Gaussian primitives at matched primitive counts. Ellipsoids consistently achieve higher PSNR and significantly faster rendering exceeding 10$\times$ those of Gaussians.
This gain stems from their compact spatial support, which reduces primitive overlap and redundant computation. Synthetic dataset experiments across dense, surface-like, and tubular geometries further confirm that ellipsoids underperform Gaussians in dense regions \sm{and in structures with diffuse boundaries}, while significantly outperforming them on surface-like and elongated structures in both PSNR and FPS (details in Appendix~\ref{app:syn_primitive}). 
\sm{This also highlights the challenge that ellipsoids' sharp boundaries present for smooth blending compared to Gaussians' soft tails.}


\label{sec:ablation-pruning}

\begin{table}[t!]
\centering
\footnotesize
\vspace{-0.5em}
\caption{
Ablation on different importance scores for structure-aware pruning. Metrics report 3D PSNR (dB) / primitive count. Results for BigBrain are reported on a single chunk.
}
\resizebox{0.9\linewidth}{!}{%
\begin{tabular}{l|cccc}
\noalign{\hrule height 1.0pt}
\textbf{Dataset} &
\textbf{w/o pruning} &
\textbf{w/o FA} &
\textbf{only FA} &
\textbf{Full} \\
\hline
{Aneurysm}
& \sm{44.91 / 32.4k}
& 49.41 / 32.3k
& 50.23 / 31.8k
& \textbf{50.76} / 32.3k \\
{Breast}
& \sm{46.43 / 96.4k}
& 45.46 / 95.3k
& 45.77 / 95.0k
& \textbf{47.34} / 95.3k \\
{BigBrain}
& \sm{16.66 / 32.9k}
& 18.69 / 33.0k
& 18.87 / 32.8k
& \textbf{18.92} / 33.0k \\
\noalign{\hrule height 1.0pt}
\end{tabular}
}
\label{tab:loss_ablation}
\end{table}







\begin{table}[t!]
\centering
\footnotesize
\renewcommand{\arraystretch}{1.0}
\setlength{\tabcolsep}{5pt}
\vspace{-0.015in}
\caption{Ablation on GPU intra-ray parallelism. Comparison of FPS before and after applying intra-ray binning to per-primitive ray sampling.}
\resizebox{\linewidth}{!}{
\begin{tabular}{l|cccc}
\noalign{\hrule height 1.0pt}
\textbf{Method} & \textbf{Aneurysm (59k)} & \textbf{Neuron (65k)} & \textbf{BigBrain (176k)} & \textbf{Hemibrain (vDeltaA) (99k)} \\
\hline
Single & 133.95 & \textbf{237.46} & 41.27 & 144.862 \\
Intra-ray (\textbf{Ours}) & \textbf{143.79} & 231.803 & \textbf{42.37} & \textbf{177.385} \\
\noalign{\hrule height 1.0pt}
\end{tabular}
}
\label{tab:intra_ray_fps}
\end{table}


\begin{figure}[h!]
 \centering
 \includegraphics[width=0.85\linewidth]{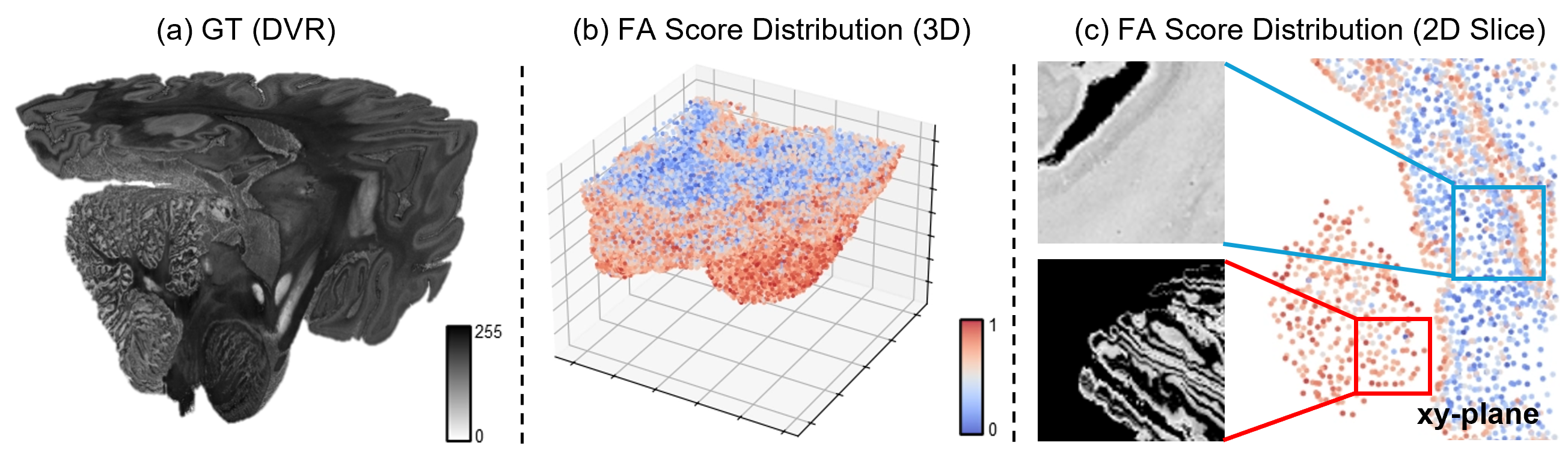}
 \vspace{-0.13in}
   \caption{
 Visualization of FA scores of ellipsoid primitives learned on a BigBrain chunk using our method. Red value indicates high FA score.
   }
 \label{fig:fa_vis}
\end{figure}

\noindent\textbf{Pruning strategy.}
Table~\ref{tab:loss_ablation} presents an ablation study of structure-aware pruning across three datasets with distinct sparsity characteristics.
\sm{As shown in the first column, primitives trained to represent the volume through densification alone yield worse reconstruction than our ADC pipeline, confirming that pruning redundant primitives after sufficient densification is more effective than sole densification.}
When removing the FA component from the importance score, reconstruction quality consistently degrades compared to using FA alone or combining FA with intensity and scale cues. The best performance is achieved when all cues are jointly considered. 
Qualitatively, as shown in Appendix Figure~\ref{fig:importance_score}, FA-based scoring helps preserve fine vascular structures and small high-frequency details (highlighted in green), while intensity and scale cues contribute to maintaining dense regions (highlighted in red). 
%
Furthermore, we observe that ellipsoids with higher FA scores tend to cluster around high-frequency, fine-scale structures, suggesting that FA effectively identifies structurally informative regions (Fig.~\ref{fig:fa_vis}).
%
%
%
%

\noindent\textbf{Rendering optimization.}
\label{sec:intra-ray}
Table~\ref{tab:intra_ray_fps} compares intra-ray parallelism performance across four datasets. While it accelerates rendering in most cases, effectiveness varies with data characteristics: full-volume structures benefit from binning, but datasets with highly compressed z-resolution like Neuron incur merging overhead due to z-axis resolution disparity, concentrating computations in single tiles.
Additional ablations are provided in the Appendix, including primitive boundary sharpness \sm{($k$)} analysis, learnable $k$, primitive type comparison \sm{on synthetic data} and \sm{number of depth bins}.
\begin{figure}[t]
    \centering
    \includegraphics[width=0.95\linewidth]{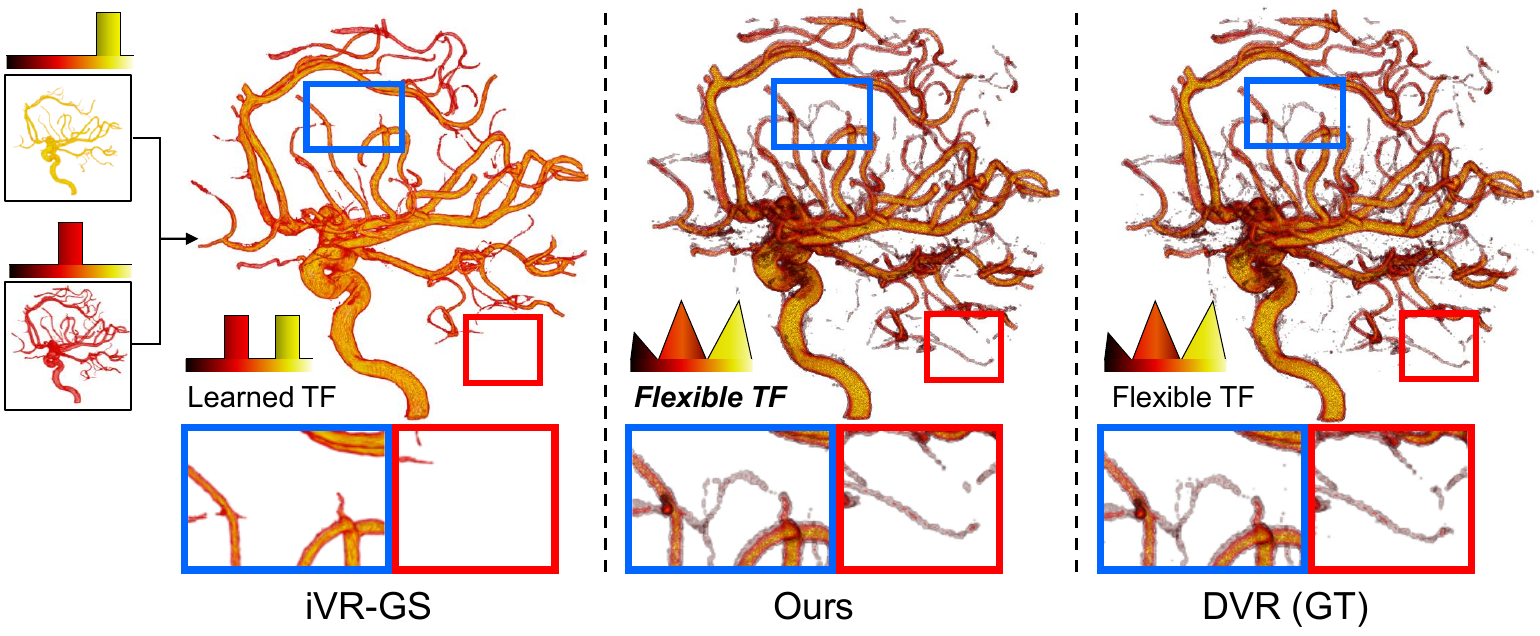}
    \vspace{-0.10in}
    \caption{
    Qualitative comparison with iVR-GS under unseen intensities.
    }
    \label{fig:ivrgs}
\end{figure}

\vspace{-0.5em}
\section{Discussion}
\label{sec:discussion}
%

Our results show that ESVR is effective for large-scale sparse volumetric data, where meaningful structures occupy only a small fraction of the domain. By placing primitives only in signal-bearing regions, ESVR achieves strong compression while preserving fine details. Compact ellipsoidal primitives and structure-aware pruning further reduce overlap and retain structures often over-smoothed by continuous neural representations. Beyond sparse data, we also evaluate dense volumes, GPU-based empty-space skipping, and mobile GPU settings to assess the method across diverse scenarios. By directly optimizing the 3D intensity field and using per-primitive ray sampling, ESVR enables accurate TF evaluation without image-based supervision or screen-space aggregation, thereby avoiding artifacts introduced by 2D rasterization-based pipelines.

\begin{figure}[!t]
    \centering
    \includegraphics[width=0.9\linewidth]{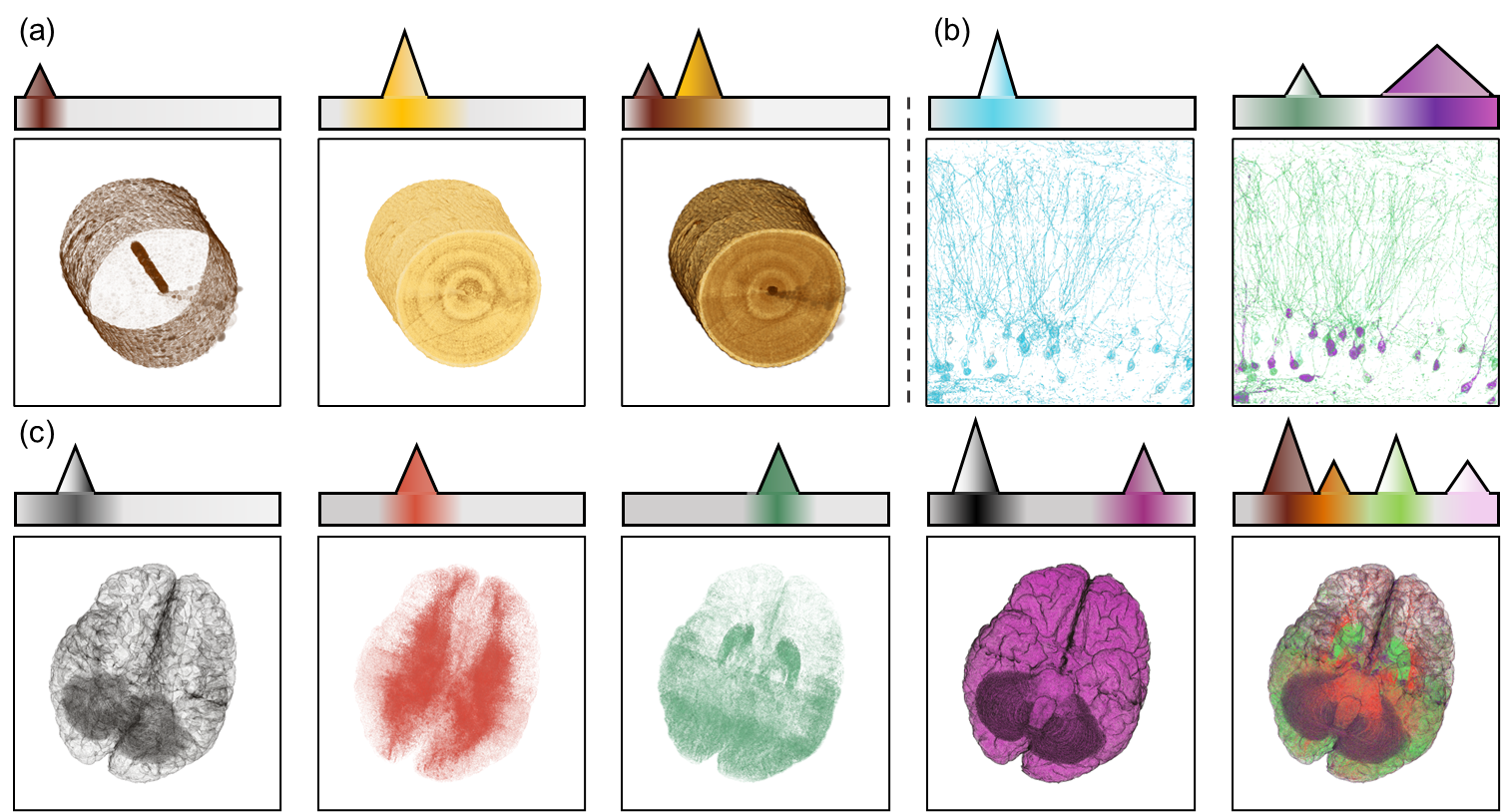}
    \vspace{-0.10in}
    \caption{Demonstration of the flexible TF capability of ESVR on three datasets : (a) Woodbranch, (b) Neuron, and (c) BigBrain.}
    \label{fig:multip_tf}
\end{figure}

\noindent\textbf{Comparison with closely related image-based methods.}
We compare our method with recent Gaussian-based approaches that share similar primitive representations but differ in optimization targets and rendering formulations.
EVER, iVR-GS, and $R^2$-Gaussian all employ Gaussian or ellipsoidal primitives to represent volumetric structures. However, they are primarily designed for image-based objectives.
\textit{EVER} performs exact ray casting using ellipsoidal primitives via explicit entry/exit computation, enabling accurate rendering at the cost of significant overhead, and is not designed for VolVis.
%
%
In particular, appearance is baked into the primitive parameters, preventing TF manipulation.
\textit{iVR-GS} is an image-based VolVis approach built on 3DGS, enabling interactive editing by decomposing appearance into palette-based parameters. However, TF control is limited to predefined palette combinations (Fig.~\ref{fig:ivrgs}), and flexibility depends on the number of trained models. 
%
\textit{$R^2$-Gaussian} incorporates volumetric information but is designed for CT reconstruction from sparse X-ray projections. It relies on projection-based supervision and total variation regularization rather than direct volumetric reconstruction loss. As a result, it achieves high 2D projection quality but lower 3D reconstruction fidelity (Appendix \sm{Table}~\ref{tab:r2}).
%
In contrast, our method directly optimizes primitives on the volumetric intensity field and supports accurate, flexible TF mapping (Fig.~\ref{fig:multip_tf}) in 3D space, enabling both compact representation and interactive visualization.

\noindent\textbf{Comparison with grid-based compression methods.}
\sm{We further compare ESVR with grid-based compression approaches, including traditional compressors (TTHRESH~\cite{ballester2019tthresh}, ZFP~\cite{lindstrom2014fixed}, and SZ3~\cite{liang2018error}) and VDB-based sparse grid representations (OpenVDB~\cite{museth2013openvdb} and NanoVDB~\cite{museth2021nanovdb}). Detailed configurations and results are provided in Appendix~\ref{app:compressor} and~\ref{app:vdb}. 
Although these methods are strong baselines for compression, their representation capacity remains tied to the underlying voxel resolution. Across both groups, we observe a common trend. Grid-based methods are highly effective on small volumes, often achieving near-lossless reconstruction or high compression ratios with fast encoding. However, as the volume resolution increases, their advantage becomes less consistent. Traditional compressors can become memory- or decompression time-bound, while lossless VDB preserves the input with substantially larger storage cost. Under aggressive size-matched compression, both traditional compressors and lossy VDB remove or smooth out sparse foreground structures, even when voxel-space PSNR remains moderate.
In contrast, ESVR allocates primitives directly to signal-bearing regions using adaptive ellipsoid primitives, making its storage less tied to the original grid resolution. This leads to a more favorable size-quality trade-off for large, sparse, and anisotropic volumes.
In addition, traditional compressors also offer significantly faster compression (typically seconds to hours versus up to days for ESVR) (Appendix Table~\ref{tab:time_comparison}). However, unlike ESVR, they require full decompression before rendering, making them less suitable for memory-constrained interactive visualization.}

\noindent\textbf{Limitations.}
Our method has several limitations.
\sm{First, while ESVR offers a rendering advantage over traditional compressors and INR-based methods, its training cost remains higher. Although chunk-based optimization enables parallelization across multiple GPUs, sequential single-GPU training of the largest volumes can take up to several days (e.g., Hemibrain~(lLN2P)), which is a non-negligible bottleneck. Promising directions to reduce this cost include skipping empty subvolumes via volumetric preprocessing and exploiting a volume prior for initialization to accelerate convergence.}
\sm{Second, the ellipsoidal primitive is inherently better suited to sparse, anisotropic structures than to dense or smoothly varying fields. On dense and diffuse-boundary data, continuous neural and Gaussian representations achieve higher reconstruction quality, as the sharp ellipsoid support cannot blend smoothly varying signals as effectively. Since a fast rendering advantage is of limited value when reconstruction quality degrades substantially, the current primitive restricts ESVR to large, sparse volumes. Designing a primitive that remains compact while faithfully representing dense regions and diffuse boundaries is a promising step toward generalizing the framework.}
Third, the method relies on several hyperparameters (e.g., primitive sharpness $k$, pruning schedule, and subvolume size) 
\sm{whose sensitivity differs sharply by data type. Sparse, high-frequency structures such as Aneurysm are highly sensitive to k (a 2.2 dB PSNR drop from k=5 to k=3, Appendix Table~\ref{app:k_ablation}), whereas dense, diffuse-boundary volumes such as BigBrain are largely k-insensitive but instead require heterogeneity-aware pruning (three pruning passes for BigBrain's cerebellar regions vs. one elsewhere, Appendix~\ref{app:imp}). No fixed setting thus generalizes across data types, motivating data-adaptive ADC and parameter-selection strategies to improve robustness and reduce manual tuning across diverse volumetric data.}
\vspace{-0.1in}
\section{Conclusion and Future Work}


We presented ESVR, an ellipsoid-based volume visualization framework that directly models volumetric data using compact primitives and per-primitive ray sampling. By fitting differentiable ellipsoids to the 3D intensity field and integrating structure-aware pruning, our method enables high compression, accurate TF evaluation, and real-time rendering.
Extensive experiments demonstrate that ESVR is particularly effective for large-scale sparse volumes, consistently achieving significantly higher rendering performance than SRN-based approaches while maintaining competitive reconstruction quality. These properties make our framework well suited for interactive and exploratory volume visualization.
Looking forward, we aim to further improve scalability and usability. Future work includes extending the framework to support out-of-core streaming, hierarchical multi-level representations, and adaptive parameter tuning. These directions will enable handling even larger datasets (e.g., terabyte-scale volumes) and broaden the applicability of primitive-based volume visualization in real-world scenarios.


\acknowledgments{
This work was supported in part by the National Research Foundation of Korea under Grant RS-2024-00349697 and Grant RS-2021-NR060143; in part by the Institute for Information and Communications Technology Planning and Evaluation under Grant IITP-2026-RS-2020-II201819; in part by the National Research Council of Science \& Technology(NST) grant by the Korea government (MSIT) (No. GTL24033-000); and in part by Korea University Grant.%
}

\bibliographystyle{abbrv-doi-hyperref}
\bibliography{template}

\appendix 
\crefalias{section}{appendix} 

\newpage
\clearpage
\setcounter{section}{0}
\setcounter{figure}{0}
\setcounter{table}{0}
\setcounter{algorithm}{0}



\section{Method Supplementary}

%
\vspace{-0.5em}
\begin{algorithm}[!htb]
\caption{Structure-aware primitive learning}
\label{alg:structure_aware_learning}
\vspace{0.2em}
\textit{$V$: ground-truth volume} \\
\textit{$E$: ellipsoid primitives with $(\mu,\Sigma,I)$} \\
\textit{$d, p$: densification and pruning iterations} \\
\textit{$\beta$: reconstruction loss weight} \\
\textit{One epoch: full traversal of subvolumes}
\begin{algorithmic}
\vspace{0.2em}
\hrule
\State $i \gets 0$ \hfill $\triangleright$ Iteration count

\While{not converged}

    \State $V_{\mathrm{gt}} \gets$ NextSubvolume($V$)
    \hfill $\triangleright$ Sequential GT subvolume

    \State $E' \gets$ GetOverlappingEllipsoids($E, V_{\mathrm{gt}}$) 
    \hfill $\triangleright$  Subvolume ellipsoids

    \State $\hat{V} \gets$ Voxelize($E'$) 
    \hfill $\triangleright$ Predicted subvolume

    \State $\mathit{L}_{\mathrm{rec}} \gets \beta \mathit{L}_1(\hat{V}, V_{\mathrm{gt}}) + (1-\beta)\mathit{L}_{\mathrm{D\mbox{-}SSIM}}(\hat{V}, V_{\mathrm{gt}})$
    \hfill $\triangleright$ Subvolume loss
    
    \State $E \gets$ Adam($\nabla \mathit{L}_{\mathrm{rec}}$)
    \hfill $\triangleright$ Update $(\mu,\Sigma,I)$

    \If{$i \in d$}
        \State Densify($E$)
        \hfill $\triangleright$ Clone / split (coverage)
    \EndIf

    \If{$i \in p$}
        \ForAll{ellipsoids $E_j \in E$}
            \State $\Omega_j \gets$ GetCoveredVoxels($E_j, V_{\mathrm{gt}}$)
            \State $\mathrm{FA}_j \gets$ ComputeFA($\Sigma_j$)
            \State $\mathrm{score}_j \gets |\Omega_j| \cdot I_j \cdot \mathrm{FA}_j$
            \hfill $\triangleright$ Importance score
        \EndFor
        \State RemoveLowScore($E$)
        \hfill $\triangleright$ Prune primitives
    \EndIf

    \State $i \gets i+1$

    \If{EndOfTraversal()}
        \State ResetSubvolumeTraversal()
        \hfill $\triangleright$ One epoch completed
    \EndIf

\EndWhile

\end{algorithmic}
\end{algorithm}

\vspace{-0.05in}
\begin{algorithm}[!hbt]
\caption{GPU-acceleration through intra-ray parallelism}
\label{alg:gpu}
\vspace{0.3em}
\textit{$w, h$: width and height of the image to rasterize} \\
\textit{$b$: number of depth bins} \\
\textit{$E$: Ellipsoid primitives with world-space parameters \\ \phantom{$E$: }($E^\mu$: mean, $E^\Sigma$: covariance)} \\
\textit{TF: transfer function to map intensity to color and opacity} \\
\textit{$View$: view configuration of current camera} \\
\textit{$I$: rendered image}

{\renewcommand{\baselinestretch}{1.01}\selectfont
\begin{algorithmic}
\vspace{0.2em}
\hrule
\vspace{0.2em}
\Function{Render}{$w, h, b, E, \textit{TF}, View$}
  \State $T \gets$ CreateTiles($w, h, b$)   
  \Comment{Tiles include depth-wise subdivisions}
  
  \State $E_{visible} \gets$ CullEllipsoid($E$, $View$) \hfill 
  
  \State $L, K \gets$ DuplicateWithKeys3D($E_{visible}, T$) \hfill
  
  \State SortByTileKeys($K, L$) \hfill 
  \Comment{No intra-tile depth sorting}

  \State $R \gets$ IdentifyTileRanges($K$)
  \Comment{range of primitives per 3D tile}
  
  \State Initialize $I_{\text{bin}} \gets 0$ 
  \Comment{Per-bin canvas}
  
  \ForAll{Tiles $t$ in $T$}
    \State $E_t \gets L[R[t]_\mathrm{start}:R[t]_\mathrm{end}]$
    \Comment{Primitives assigned to tile $t$}
    
    \ForAll{Pixels $i$ in $t$}
      \State $\text{ray}[i,t_z]\gets$ IntensitySample($i, E^{\mu}_{t},E^{\Sigma}_{t}, \textit{TF}$)
      \State $I_{\text{bin}}[i,t_z] \gets$ BlendPartialRay($\text{ray}[i,t_z]$)
    \EndFor
  \EndFor
  \State Flatten $T \gets$ CreateTiles($w, h, 1$) \hfill 
  \Comment{Reset $tile\_grid.z = 1$}
  \ForAll{Pixels $i$ in $I$}
    \State $I[i] \gets$ MergeZBins($I_{\text{bin}}[i, 0{:}b])$ \hfill 
    \Comment{Blend across $b$ bins}
  \EndFor
  \State \Return $I$
\EndFunction
\end{algorithmic}
}
\end{algorithm}

\subsection{Structure-aware Primitive Learning}
\noindent Appendix Algorithm~\ref{alg:structure_aware_learning} summarizes the full optimization loop of our structure-aware primitive learning, including 3D intensity field fitting, adaptive densification, and structure-aware pruning.
\subsection{GPU Intra-Ray Parallelism Pseudo Code}
\label{app:GPUIntraPseudo}
Appendix Algorithm~\ref{alg:gpu} describes our GPU-acceleration scheme based on intra-ray parallelism, which partitions each ray into depth bins and enables efficient parallel accumulation of primitive contributions during rendering.


\section{Method Figures}

\subsection{Differentiable Ellipsoid Primitive}
\begin{figure}[!h]
 \centering
 \includegraphics[width=0.9\linewidth]{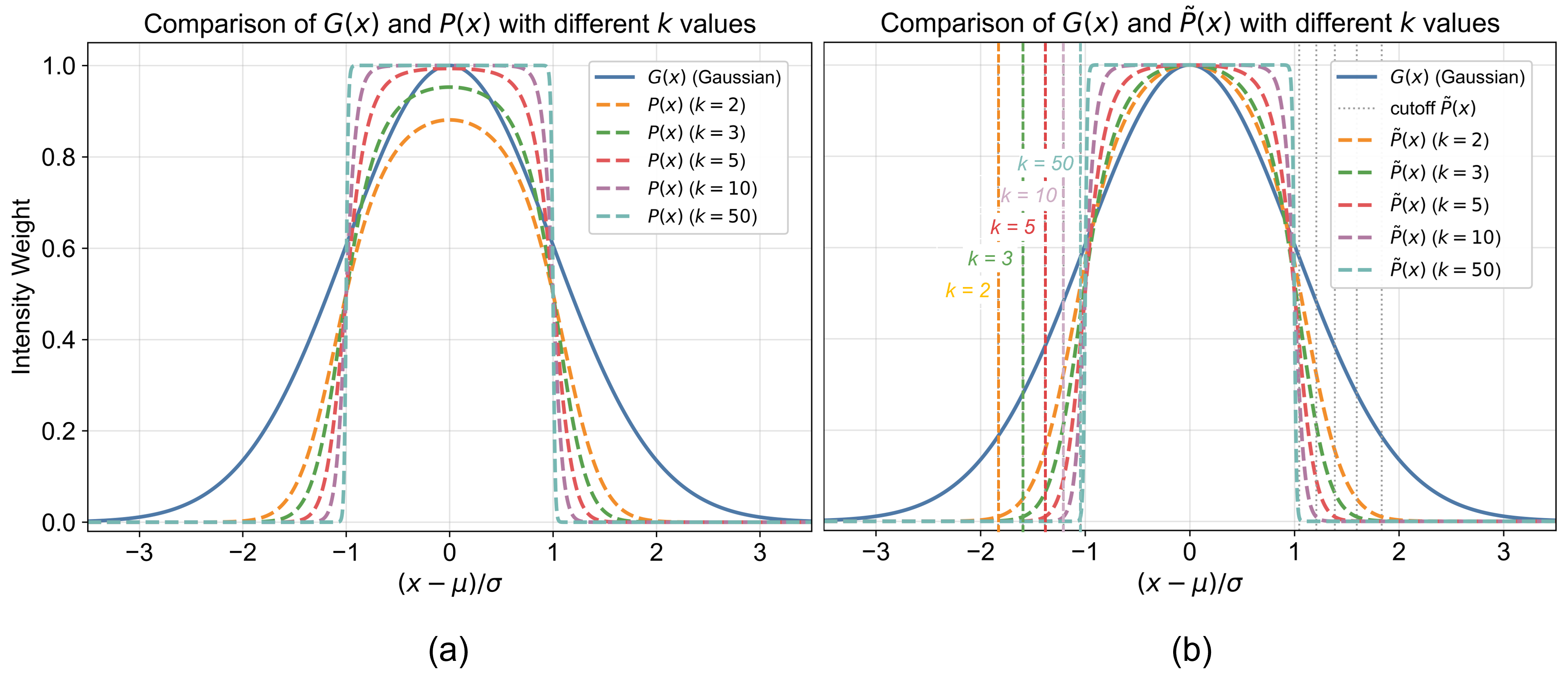}
 \vspace{-3pt}
\caption{
1D comparison between the Gaussian kernel $G(x)$ and our sigmoid-based ellipsoid primitives with unit center intensity. (a) Effect of the sharpness parameter $k$ on the primitive $P(x)$. Peak intensity decreases as $k$ becomes smaller than 5. (b) Normalized primitive $\tilde{P}(x)$ with corresponding cutoff regions \sm{for $k = 2, 3, 5, 10, 50$.}
}
 \label{fig:1d_primitives}
\end{figure}

\noindent Appendix Figure~\ref{fig:1d_primitives} compares the 1D spatial support of a Gaussian kernel and our sigmoid-based ellipsoidal primitive. As shown in Appendix Figure ~\ref{fig:1d_primitives} (a), the unnormalized primitive ${P}(x)$ exhibits reduced peak intensity as the sharpness parameter $k$ decreases, deviating from the desired unit response at the center. To address this, we normalize the primitive $\tilde{P}(x)$, as shown in Appendix Figure ~\ref{fig:1d_primitives} (b), ensuring that the peak intensity remains constant regardless of $k$. In addition, the spatial support varies with $k$, and the cutoff region is adjusted accordingly to avoid truncation artifacts.

\subsection{ESVR Interactive Viewer}
\begin{figure}[!h]
 \centering
 \includegraphics[width=0.8\linewidth]{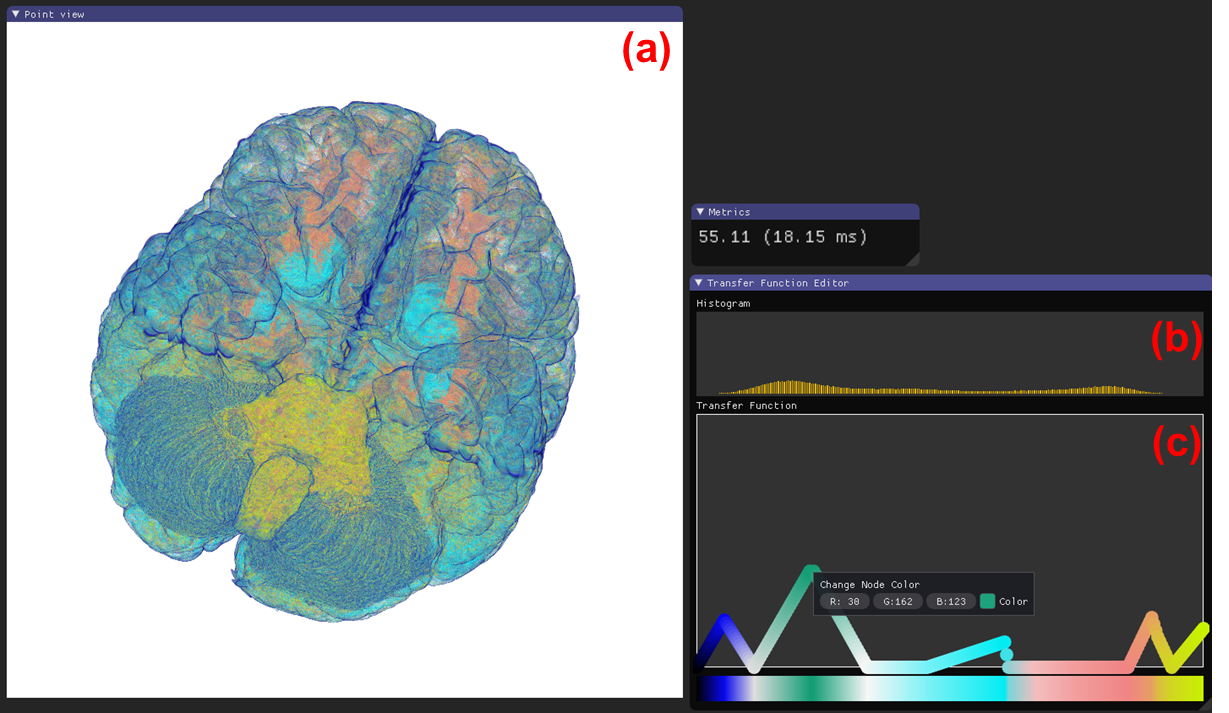}
   \caption{TF editing interface of the ESVR viewer. (a) Rendering result of the extended CUDA-based rasterizer. 
(b) Histogram of data intensities. 
(c) TF editor for mapping intensities to color and opacity.}
 \label{fig:tf_viewer}
\end{figure}

\noindent Appendix Figure~\ref{fig:tf_viewer} presents the proposed viewer for flexible TF editing and interactive volume visualization. A demonstration of user interaction is provided in the supplemental video.


\section{Experiment Configuration}
\begin{table}[!htb]
\caption{Training hyperparameters of ESVR}
\vspace{-1pt}
\centering
\scriptsize
\setlength{\tabcolsep}{3pt}
\renewcommand{\arraystretch}{1.0}

\resizebox{\linewidth}{!}{%
\begin{tabular}{l|ccccccccc}
\noalign{\hrule height 1.0pt}
\textbf{Dataset}
& $n_{\mathrm{chunk}}$ & $res_{\mathrm{chunk}}$ & $res_{\mathrm{subvol}}$
& $e_{\mathrm{total}}$ 
& $e_{\mathrm{densify}}$
& $e_{\mathrm{prune}}$
& $thres_{\mathrm{grad}}$
& $per_{\mathrm{dense}}$
\\
\noalign{\hrule height 1.0pt}

Aneurysm 
& - & - & $128^3$ & 2800 
& 16/400 & 1200 
& $2.5\mathrm{e}{-9}$ 
& 0.008 
\\

Breast 
& - & - & $128^3$ & 2800 
& 16/400 & 1200 
& $2.5\mathrm{e}{-9}$ 
& 0.008 
\\

Neuron
& - & - & $133^3$ & 2800 
& 16/400 & 1200 
& $6.5\mathrm{e}{-9}$ 
& 0.001 
\\

BigBrain 
& 8 & (875,822,714) & $440^3$ & 1600 
& 16/400 & 200/300/800 
& $2.5\mathrm{e}{-9}$ 
& 0.008 
\\

Woodbranch 
& 8 & (1024,1024,1024) & $512^3$ & 1600 
& 16/400 & 800 
& $2.5\mathrm{e}{-9}$ 
& 0.01 
\\

Hemibrain(vDeltaA) 
& 64 & (1093,1018,1043) & $550^3$ & 800 
& 16/200 & 400 
& $1.0\mathrm{e}{-10}$ 
& 0.01 
\\

Hemibrain(lLN2P) 
& 125 & (1226,955,1023) & $614^3$ & 800 
& 16/200 & 400 
& $1.0\mathrm{e}{-10}$ 
& 0.01 
\\

\noalign{\hrule height 1.0pt}
\end{tabular}
}

\label{tab:hyperparameter_ours}
\end{table}
\begin{table*}[!htb]
\centering
\caption{Training hyperparameters for baseline methods. For large datasets, we report a per-chunk network configuration.}
\small
\setlength{\tabcolsep}{4pt}
\renewcommand{\arraystretch}{1.0}

\begin{tabular}{ll|cc|cccc|ccc|ccc}
\noalign{\hrule height 1.0pt}

& 
& \multicolumn{2}{c|}{\textbf{fV-SRN}}
& \multicolumn{4}{c|}{\textbf{NGP}}
& \multicolumn{3}{c|}{\textbf{AMGSRN++}}
& \multicolumn{3}{c}{\textbf{InstantVNR}} \\

\textbf{Dataset} & \textbf{Config}
& $n_{\mathrm{feat}}$ & $res_{\mathrm{grid}}$
& $n_{\mathrm{lev}}$ & $\log_2 h$ & $res_{\mathrm{base}}$ & $res_{\mathrm{max}}$
& $n_{\mathrm{feat}}$ & $n_{\mathrm{grid}}$ & $res_{\mathrm{grid}}$
& $n_{\mathrm{lev}}$ & $\log_2 h$ & $res_{\mathrm{base}}$ \\

\noalign{\hrule height 1.0pt}

\multirow{2}{*}{Aneurysm}
& Size-M  
& 2  & (72,72,72) 
& 14 & 15 & 16 & 512
& 2  & 10 & (32,32,32)
& 12 & 15 & 16\\
& PSNR-M  
& 8 & (128,128,128)
& 17& 20& 16 & 256
& 4& 16& (64,64,64)
& 19& 19& 16\\
\hline

\multirow{2}{*}{Breast}
& Size-M  
& 2  & (72,72,72) 
& 14 & 15 & 16 & 512
& 2  & 12 & (32,32,32)
& 13 & 15 & 16\\
& PSNR-M  
& 8& (256,256,256) 
& 17& 24& 16& 256
& 4& 16& (128,128,128)
& 16& 18& 16\\
\hline

\multirow{2}{*}{Neuron}
& Size-M  
& 2& (75,75,75)
& 14& 15& 16& 512
& 2& 12& (32,32,32)
& 9& 15& 16\\
& PSNR-M  
& 4& (64,64,64) 
& 17& 15& 16& 512
& 2& 12& (101,52,6) 
& 16& 17& 16\\
\hline

\multirow{2}{*}{BigBrain}
& Size-M  
& 2& (49,49,49)
& 16& 13& 16& 512
& 2& 14& (22,22,22)
& 12& 16& 16\\
& PSNR-M  
& 2& (50,50,50)
& 10& 13& 16& 512
& 2& 12& (20,20,20)
& 14& 17& 16\\
\hline

\multirow{2}{*}{Woodbranch}
& Size-M  
& 4& (32,32,32) 
& 17& 12& 16& 512
& 2& 16& (16,16,16)
& 10& 15& 16\\
& PSNR-M  
& 4&(30,30,30) 
& 15& 12& 16& 512
& 2& 12& (16,16,16)
& 14& 17& 16\\
\hline

\multirow{2}{*}{Hemibrain (vDeltaA)}
& Size-M  
& 2& (16,16,16)
& 9& 10& 8& 512
& 1& 10& (10,10,10)
& -& -& -\\
& PSNR-M  
& 4& (24,24,24)
& 10& 10& 8& 512
& 1& 16& (10,10,10)
& -& -& -\\
\hline

\multirow{2}{*}{Hemibrain (lLN2P)}
& Size-M  
& 2& (20,20,20)
& 9& 10& 8& 512
& 1& 16& (10,10,10)
& 14& 16& 16\\
& PSNR-M  
& 4& (24,24,24)
& 8& 10& 8& 256
& 1& 16& (10,10,10)
& 14& 16& 16\\
\hline

\noalign{\hrule height 1.0pt}
\end{tabular}

\label{tab:baseline_config}
\end{table*}

\newpage
\subsection{Dataset Details}
\label{app:dataset-details}
The hemibrain dataset \cite{scheffer2020connectome} is a synapse-resolution connectome of the adult Drosophila melanogaster brain, containing $\sim$ 25,000 neurons and millions of synaptic connections. Volumes corresponding to specific annotated neuron types (e.g., vDeltaA and lLN2P) are extracted using neuPrint \cite{plaza2022neuprint} for querying neuron metadata and CloudVolume \cite{cloudvolume} for retrieving volumetric data. Due to the large size of full-resolution volumes, we use downsampled representations at MIP level 2 ($\times$8 downsampling) or 3 ($\times$64 downsampling) for analysis.
The BigBrain dataset \cite{amunts2013bigbrain} is a 3D reconstruction of a complete human brain at 20$\mu\mathrm{m}$ isotropic resolution. Similar to the hemibrain dataset, we use downsampled volumes at MIP level 2.
In addition, we utilize a Neuron dataset, a proprietary dataset provided by collaborators.

\subsection{Training Implementation Details}
\label{app:imp}
Appendix Table \ref{tab:hyperparameter_ours} summarizes the hyperparameters used for training, including number of chunks ($n_{\mathrm{chunk}}$), chunk resolution ($res_{\mathrm{chunk}}$), subvolume resolution ($res_{\mathrm{subvol}}$), total epochs ($e_{\mathrm{total}}$), densification start/end epoch ($e_{\mathrm{densify}}$), structure-aware pruning epoch ($e_{\mathrm{prune}}$), gradient threshold ($thres_{\mathrm{grad}}$) and percent-dense parameter ($per_{\mathrm{dense}}$) for densification. Note that per-chunk parameters are denoted for large volumes. 
\label{app:hyperparm}

\noindent\textbf{Chunk-based training.}
Chunk-based training is applied to large-scale datasets with different numbers of chunks: using 8 chunks for BigBrain and Woodbranch, 64 for Hemibrain (vDeltaA), and 125 for Hemibrain (lLN2P). 
For all datasets, the full volume is optimized in a normalized $\text{[-1,1]}^3$ coordinate space. For small and medium volumes, ellipsoids are trained across the entire $\text{[-1,1]}^3$ space, while the training space for large volumes is scaled accordingly (e.g., $\text{[-0.5,0.5]}^3$ for 8 chunks, $\text{[-0.25,0.25]}^3$ for 64 chunks, and  $\text{[-0.2,0.2]}^3$ for 125 chunks).
\sm{Regarding hardware, all training is performed on NVIDIA RTX A6000 GPUs. BigBrain and Woodbranch are trained sequentially on a single A6000, while the two large Hemibrain datasets use the parallel chunk-based mode across
multiple GPUs: 4 GPUs for Hemibrain~(vDeltaA) and 5 GPUs for Hemibrain~(lLN2P). All small and medium datasets are trained on a single A6000.}

\noindent\textbf{Subvolume configuration.}
During optimization, the training unit (either the full volume or an individual chunk) is partitioned into voxelized cubic subvolumes. The subvolume side length is set to approximately half the longest axis of the training unit. For flat volumes such as Neuron, the smallest axis is used instead.

\noindent\textbf{Training schedule.}
Optimization begins with a 16 epoch warm-up phase without structural updates, followed by densification every 4 epochs. Along with 3D intensity field fitting, densification is performed until 14--25$\%$ of the training process, while structure-aware importance score pruning is applied at 42--50$\%$. The total number of training epochs is scaled with volume size (2800 / 1600 / 800 for small–medium / large / extremely large volumes) to balance training time and quality.
All datasets execute single pruning iteration except for BigBrain, which is treated as an exception with a multi-stage pruning strategy. The BigBrain dataset contains heterogeneous regions, including relatively smooth regions (cortical) and highly detailed regions (cerebellar), which require more aggressive densification. 
For the cerebellar chunks, pruning is performed three times: twice during the densification phase at epochs 200 and 300 (pruning ratio 0.8), and once after densification at epoch 800 (ratio 0.85).

\noindent\textbf{Densification and gradient threshold.}
Densification follows the original 3DGS principle: primitives with large gradients indicate regions requiring more representational capacity and are selected for cloning or splitting. This gradient-based selection ensures efficient allocation of primitives to regions with high reconstruction error.
The densification process is controlled by two key parameters. One is the gradient threshold ($thres_{\mathrm{grad}}$), determining whether a primitive should be densified based on the magnitude of its gradient, while the other is the percent-dense parameter ($per_{\mathrm{dense}}$), determining whether to split or clone the primitive based on the spatial extent. 
Each parameter is tuned depending on the dataset and volume resolution:
the gradient threshold ($1 \times 10^{-10}$ to $6.5 \times 10^{-9}$) and the percent-dense parameter ($0.001$ to $0.01$).
These parameters jointly control the aggressiveness of densification.

\subsection{Baseline Configuration}
Appendix Table \ref{tab:baseline_config} summarizes the hyperparameters used for baseline SRN methods. All parameters for each method are adjusted to match the target model size or PSNR. Parameters not listed in the table follow the default settings of each method.

\noindent\textbf{(1) fV-SRN~\cite{weiss2022fast}} employs latent grids as its primary representation, with the number of features per grid ($n_{\mathrm{feat}}$) and grid resolution ($res_{\mathrm{grid}}$) being adjustable. 

\noindent\textbf{(2) NGP~\cite{muller2022instant}} leverages multi-resolution hash encoding for fast training. The key parameters that determine the hash table capacity are the number of levels  ($n_{\mathrm{lev}}$), hash table size ($\log_2 h$), and base/max grid resolution ($res_{\mathrm{base}}, res_{\mathrm{max}}$).

\noindent\textbf{(3) AMGSRN++~\cite{wurster2025amgsrn++}} is based on the prior work~\cite{wurster2023adaptively} which introduces an adaptive multi-grid representation for scientific visualization. AMGSRN++ further enhances this framework by incorporating custom CUDA kernels, feature grid compression, and extensions to the temporal domain.
Encoding is mainly configured using three parameters: the number of features per grid ($n_{\mathrm{feat}}$), the number of grids ($n_{\mathrm{grid}}$), and the grid resolution ($res_{\mathrm{grid}}$).

\noindent\textbf{(4) InstantVNR~\cite{wu2023interactive}} is a volume rendering framework based on NGP ~\cite{muller2022instant}, extended with sample streaming, macro-cell optimization, and out-of-core sampling techniques. Although it adopts the same encoding scheme as NGP, the network parameters are not exactly identical, as the macro-cell structure also contributes to the overall model size when matching the target model size or PSNR.

\subsection{iVR-GS Training Details}
iVR-GS \cite{tang2025ivr} composes multiple base models where each model is trained under one basic TF. For each basic scene, Gaussians are optimized under a view-dependent appearance of the rendered scene using a Blinn-Phong shading model. In this formulation, each Gaussian's color is decomposed into intrinsic color and lighting components, enabling post-training editing of both appearance and illumination. 
This design differs from our experimental setup, which assumes a lighting-free (emission-based) volumetric representation. 
To ensure identical rendering settings with our method, the training images for iVR-GS are generated from DVR images with lighting disabled.
Otherwise, the inclusion of shading effects in the input images would lead to appearance differences.
We train two basic models for Aneurysm dataset using 180 multi-view images and evaluate them on 180 novel views, following the icosphere sampling strategy of the original paper. 
For visualization, the iVR-GS result presented in Figure \ref{fig:ivrgs} in the main text is rendered using its default viewer. 
Although we disabled lighting in the training images, minor brightness variations may still persist due to the view-dependent nature of the iVR-GS training formulation.
However, as our primary goal is to evaluate the information loss of image-based methods, we focus the comparison on structural differences rather than appearance variations.

\section{Additional Experiments}

\subsection{Comparison with Image-based Baseline Methods}
\noindent\textbf{${R}^2$-Gaussian.}
Appendix Table~\ref{tab:r2} presents a quantitative comparison with the baseline, ${R}^2$-Gaussian \cite{zha2024r}. 
${R}^2$-Gaussian reconstructs volumes from X-ray projections. To ensure a fair comparison, we synthesize X-ray projections from the ground-truth volume using TIGRE\footnotemark[1] and provide both the synthetic projections and the ground-truth volume to ${R}^2$-Gaussian during reconstruction. 
As ${R}^2$-Gaussian is trained primarily from X-ray projections while using volumetric data only as a regularization signal, its reconstructed volumes exhibit significantly lower fidelity in 3D space, as reflected by substantially lower 3D PSNR values across all datasets (e.g., 22.92 vs. 51.54 on the Aneurysm dataset). In contrast, ${R}^2$-Gaussian achieves relatively higher 2D PSNR, since it is optimized to match X-ray projections. However, this metric reflects projection consistency rather than volumetric reconstruction quality, and thus cannot be directly applied to volume visualization tasks. Finally, due to its image-based training paradigm, ${R}^2$-Gaussian demonstrates notably faster training times compared to our method. Due to the lack of support for large raw volumetric data in ${R}^2$-Gaussian, comparisons are limited to datasets up to the Neuron scale.

\begin{table}[H]
\centering
\caption{ Quantitative comparison with $R^2$-Gaussian. Training time is measured from initialization to completion. Values in parentheses denote the number of primitives for each method.}
\small
\setlength{\tabcolsep}{6pt}
\renewcommand{\arraystretch}{1.0}
\begin{tabular}{l|c|ccc}
\noalign{\hrule height 1.0pt}
\multirow{2}{*}{\textbf{Dataset}}  & \textbf{Ours} & \multicolumn{3}{c}{\textbf{$R^2$-Gaussian}} \\
 &  \textbf{3D PSNR} & \textbf{3D PSNR} & \textbf{2D PSNR} & \textbf{Training Time (s)} \\
\hline
Aneurysm (60.4k) & 51.54 (59k) & 22.92 & 53.27 & 553 \\
Breast (69.2k)  & 46.00 (67k) & 21.01 & 54.27 & 616 \\
Neuron (65.3k)  & 39.56 (65k) & 31.86 & 47.78 & 1063 \\
\noalign{\hrule height 1.0pt}
\end{tabular}
\label{tab:r2}
\end{table}
%



\noindent\textbf{LightGaussian.}
\label{app:lightgaussian}
Appendix Table~\ref{tab:lightgaussian} summarizes the pruning performance comparison with LightGaussian~\cite{fan2024lightgaussian}. 
Across three datasets of varying scale, our method achieves higher PSNR and SSIM while using a comparable or smaller number of primitives. LightGaussian scores primitives based on accumulated ray contributions from multiple synthesized views in a 2D projection-based framework, using view-dependent features such as ray hits and transmittance. Following this formulation, we generate 360 uniformly sampled views on a sphere for our experimental setting. When applied to our volume setting, this projection-based scoring yields lower accuracy (for example, 48.76 dB on Aneurysm and 38.34 dB on Neuron), compared to our volume-based scoring (50.76 dB and 39.44 dB). In contrast, our method computes importance directly in volumetric space by integrating voxel intensities covered by each primitive and weighting them with fractional anisotropy, leading to more reliable ranking of primitives.
\footnotetext[1]{A. Biguri et al.,  Tigre: a matlab-gpu toolbox for cbct image reconstruction. \textit{Biomedical Physics \& Engineering Express}, 2(5):055010, 2016.}


\begin{table}[H]
\centering
\small
\renewcommand{\arraystretch}{1.0}
\setlength{\tabcolsep}{3pt}
\caption{
Pruning performance comparison with LightGaussian. Metrics report 3D PSNR (\,dB\,) / SSIM / number of primitives. For Woodbranch we use one of the chunks for evaluation.}
\resizebox{1.0\linewidth}{!}{%
\begin{tabular}{l|ccc}
\noalign{\hrule height 1.0pt}
\textbf{Method} & \textbf{Aneurysm} & \textbf{Neuron} & \textbf{Woodbranch} \\
\hline
LightGaussian & 48.76 / 0.9968 / 32k & 38.34 / 0.9206 / 77k & 41.19 / 0.9147 / 12k \\
\textbf{Ours} & \textbf{50.76} / 0.9976 / 32k & \textbf{39.44} / 0.9267 / 77k & \textbf{41.04} / 0.8808 / 12k \\
\noalign{\hrule height 1.0pt}
\end{tabular}%
}
\label{tab:lightgaussian}
\end{table}

\subsection{Comparison with Traditional Compression Methods}
\label{app:compressor}

\noindent Though data reduction is not the primary objective of our method, we believe it is valuable to compare the compression capability of ESVR with traditional compression techniques. We emphasize that our method is not a compressor, but a volumetric representation and rendering framework.
%
We compare against three traditional compression methods, TTHRESH~\cite{ballester2019tthresh}, ZFP~\cite{lindstrom2014fixed}, and SZ3~\cite{liang2018error}. 
We do not include rendering-based compressors or bricked variants of existing methods, as prior work~\cite{weiss2022fast} has shown that such approaches typically result in slower decoding, lower compression ratios, and higher memory usage. 
%
Our evaluation metrics include compressed size, 3D PSNR, compression time, and decompression time. Notably, for our method, compression time represents the full training time, and decompression time is the time required to query the entire volume via 3D rasterization (i.e., voxelization).
All compression methods are evaluated on a system with dual Intel Xeon Gold 6226R CPUs (32 cores, 64 threads), 1.1~TB RAM, and an NVIDIA RTX A6000 GPU. We use CPU-based implementations for all traditional compressors to ensure a consistent evaluation environment. 
Although ZFP provides GPU-accelerated variants, they are strictly limited to fixed-rate mode and do not support the flexible error-bounded configurations required for our study.

%
%
\noindent\textbf{Compression Results.}
\sm{
Appendix Table~\ref{tab:compression-main} shows the quantitative results under PSNR-matched and Size-matched settings, where compression methods are configured to match either the PSNR levels or memory size reported in Table~\ref{tab:volumetric_comparison} of main text. 
}
\sm{
The relative performance depends strongly on dataset scale. 
Under the 3D PSNR-matched setting, traditional compressors achieve substantially smaller storage sizes and nearly lossless reconstruction for Aneurysm.
However, as the dataset size increases, ESVR increasingly outperforms ZFP and SZ3 in compression ratio. TTHRESH remains highly effective on Neuron and Woodbranch, achieving even smaller model sizes than ESVR, but fails earlier with out-of-memory errors on Hemibrain due to its higher memory demand during compression.
The Size-matched setting shows a similar trend. On small datasets, traditional compressors can maintain high quality and sometimes undershoot the target size. On larger datasets, however, matching ESVR's compact size requires aggressive voxel-level compression for ZFP and SZ3, leading to lower PSNR.
TTHRESH, in contrast, achieves higher PSNR than ESVR on Neuron and Woodbranch under the same size budget, while encountering the same OOM failure on Hemibrain.
\begin{table}[htb]
\centering
\caption{
\sm{Comparison with traditional compression methods under 3D PSNR-matched and Size-matched settings.}
}
\footnotesize
\label{tab:compression-main}
\setlength{\tabcolsep}{4.0pt}
\renewcommand{\arraystretch}{0.95}
\resizebox{1.0\columnwidth}{!}{%
\begin{tabular}{ll|c |ccc| ccc}
\noalign{\hrule height 1.0pt}
\multirow{2}{*}{Dataset}
& \multirow{2}{*}{Metric}
& \multirow{2}{*}{Ours}
& \multicolumn{3}{c}{\rule{0pt}{2.2ex}3D PSNR-matched}
& \multicolumn{3}{c}{\rule{0pt}{2.2ex}Size-matched} \\
\cline{4-6}
\cline{7-9}
& & 
& \rule{0pt}{2.2ex}TTHRESH & ZFP & SZ3
& \rule{0pt}{2.2ex}TTHRESH & ZFP & SZ3 \\
\hline

\multirow{4}{*}{Aneurysm}
& \rule{0pt}{2.2ex}Size (MB)
& 2.600 & 3.318 & 0.301 & 0.154
& 2.611 & 1.805 & 0.320 \\
& 3D PSNR (dB)
& 51.54 & 51.56 & 54.42 & 51.28
& 47.74 & 183.60 & $\infty$ \\
& Comp. (s)
& 388 & 23.7459 & 0.1457 & 0.5319
& 26.27 & 0.15 & 0.07 \\
& Decomp. (s)
& 0.0079 & 1.7490 & 0.1473 & 0.1343
& 1.41 & 0.08 & 0.06 \\
\hline

\multirow{4}{*}{Neuron}
& \rule{0pt}{2.2ex}Size (MB)
& 2.800 & 1.175 & 11.567 & 2.060
& 2.866 & 3.417 & 2.795 \\
& 3D PSNR (dB)
& 39.42 & 39.49 & 40.15 & 39.57
& 40.44 & 34.17 & 40.41 \\
& Comp. (s)
& 1320 & 159.9387 & 4.0231 & 17.1053
& 330.33 & 2.49 & 11.96 \\
& Decomp. (s)
& 0.0482 & 25.6487 & 3.4315 & 3.2295
& 38.87 & 2.32 & 5.55 \\
\hline

\multirow{4}{*}{Woodbranch}
& \rule{0pt}{2.2ex}Size (MB)
& 4.300 & 0.188 & 239.558 & 31.090
& 4.170 & 16.777 & 4.355 \\
& 3D PSNR (dB)
& 40.60 & 40.69 & 40.94 & 40.60
& 42.18 & 28.41 & 39.35 \\
& Comp. (s)
& 5676 & 3406.1630 & 149.1826 & 355.3101
& 6779.67 & 37.68 & 232.69 \\
& Decomp. (s)
& 1.0931 & 587.1047 & 109.1363 & 106.1914
& 760.76 & 36.62 & 106.58 \\
\hline

\multirow{4}{*}{\shortstack{Hemibrain\\(vDeltaA)}}
& \rule{0pt}{2.2ex}Size (MB)
& 4.200 & \textsc{oom} & 147.590 & 5.169
& \textsc{oom} & 145.065 & 4.194 \\
& 3D PSNR (dB)
& 35.34 & \textsc{oom} & 35.86 & 35.57
& \textsc{oom} & 30.06 & 32.68 \\
& Comp. (s)
& 2998 & \textsc{oom} & 240.4151 & 4493.3396
& \textsc{oom} & 308.16 & 4095.52 \\
& Decomp. (s)
& 5.1207 & \textsc{oom} & 285.3800 & 1795.7484
& \textsc{oom} & 1964.00 & 2505.96 \\
\noalign{\hrule height 1.0pt}
\end{tabular}%
}
\end{table}

\begin{figure}[htb]
    \centering
    \includegraphics[width=1.0\linewidth]{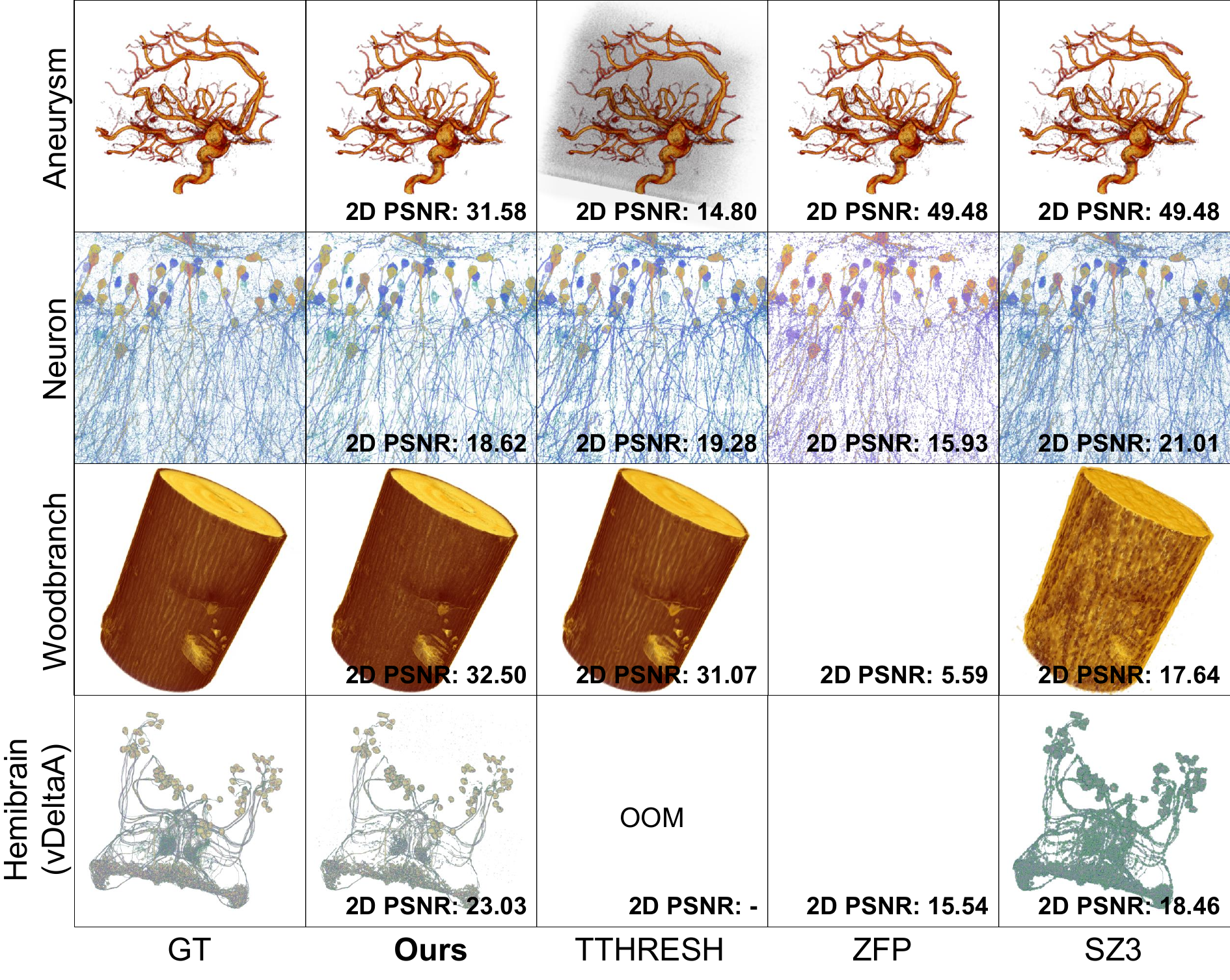}
    \vspace{-0.5em}
    \caption{\sm{Qualitative results of traditional compression methods. Woodbranch and Hemibrain are downsampled (×8 and ×64) to enable DVR (GT, TTHRESH, ZFP, SZ3) rendering without OOM.}}
    \label{fig:size-match-rendering}
\end{figure}
}
%

\noindent\textbf{Qualitative Comparison.}
\sm{
\sm{Although 3D PSNR captures the overall reconstruction trend, it can overestimate visual fidelity for sparse volumes because most voxels correspond to background or empty space. 
\begin{table}[!h]
\centering
\small
\caption{\sm{Time comparison for compression/decompression and single-frame rendering. For ours, compression time indicates sequential training time on a single GPU. Rendering time is reported as seconds per frame. Total time is the sum of the above, denoting the time required to render a single frame.}}
\label{tab:time_comparison}
\vspace{-0.5em}
\begin{tabular}{ll|c|c|c|c}
\noalign{\hrule height 1.0pt}
Dataset & Metric & Ours & TTHRESH & ZFP & SZ3 \\
\hline
\multirow{4}{*}{Aneurysm}
& Comp. (s) & 388 & 23.75 & 0.15 & 0.53 \\
& Decomp. (s) & - & 1.75 & 0.15 & 0.13 \\
& Render (s) & 0.0045 & 0.0073 & 0.0073 & 0.0073 \\
& Total (s) & 388.00 & 25.51 & 0.31 & 0.67 \\
\hline
\multirow{4}{*}{Breast}
& Comp. (s) & 756 & 11.08 & 0.23 & 1.23 \\
& Decomp. (s) & - & 2.55 & 0.23 & 0.25 \\
& Render (s) & 0.0047 & 0.0047 & 0.0047 & 0.0047 \\
& Total (s) & 756.00 & 13.63 & 0.46 & 1.48 \\
\hline
\multirow{4}{*}{Neuron}
& Comp. (s) & 1320 & 159.94 & 4.02 & 17.11 \\
& Decomp. (s) & - & 25.65 & 3.43 & 3.23 \\
& Render (s) & 0.0076 & 0.0110 & 0.0110 & 0.0110 \\
& Total (s) & 1320.01 & 185.60 & 7.46 & 20.35 \\
\hline
\multirow{4}{*}{BigBrain}
& Comp. (s) & 32320 & 4233.14 & 44.57 & 134.62 \\
& Decomp. (s) & 0.6731 & 714.90 & 38.98 & 36.46 \\
& Render (s) & 0.0206 & 0.0693 & 0.0693 & 0.0693 \\
& Total (s) & 32320.02 & 4948.11 & 83.62 & 171.15 \\
\hline
\multirow{4}{*}{Woodbranch}
& Comp. (s) & 45408 & 3406.16 & 149.18 & 355.31 \\
& Decomp. (s) & - & 587.10 & 109.14 & 106.19 \\
& Render (s) & 0.0230 & \textsc{oom} & \textsc{oom} & \textsc{oom} \\
& Total (s) & 45408.02 & - & - & - \\
\hline
\multirow{4}{*}{\makecell{Hemibrain \\ (vDeltaA)}}
& Comp. (s) & 191872 & \textsc{oom} & 240.42 & 4493.34 \\
& Decomp. (s) & - & \textsc{oom} & 285.38 & 1795.75 \\
& Render (s) & 0.0072 & \textsc{oom} & \textsc{oom} & \textsc{oom} \\
& Total (s) & 191872.01 & - & - & - \\
\hline
\multirow{4}{*}{\makecell{Hemibrain \\ (lLN2P)}}
& Comp. (s) & 440250 & \textsc{oom} & 501.98 & \textsc{oom} \\
& Decomp. (s) & - & \textsc{oom} & 541.65 & \textsc{oom} \\
& Render (s) & 0.0115 & \textsc{oom} & \textsc{oom} & \textsc{oom} \\
& Total (s) & 440250.01 & - & - & - \\
\noalign{\hrule height 1.0pt}
\end{tabular}
\end{table}
\noindent
Therefore, rendering quality is also an important evaluation criterion for sparse volumetric data. As shown in Appendix Figure~\ref{fig:size-match-rendering}, ZFP reconstructs the small Aneurysm dataset nearly losslessly at a substantially smaller size, but its renderings become visibly blocky from Neuron onward and fail to preserve foreground structures on Woodbranch and Hemibrain. This discrepancy is more clearly reflected by 2D PSNR. For example, ZFP achieves only 15.93, 5.59, and 15.54 dB on Neuron, Woodbranch, and Hemibrain, respectively, compared with 18.62, 32.50, and 23.03 dB for our method. SZ3 remains competitive on Neuron, but its rendered results on large datasets deviate visibly from the ground truth, consistent with its substantially lower 2D PSNR on Woodbranch and Hemibrain. TTHRESH achieves slightly higher 2D PSNR on Neuron, whereas our method is better on Woodbranch. On Hemibrain, TTHRESH runs out of memory. Taken together, these results highlight a broader limitation of regular-grid compression for sparse volumetric data. Under aggressive memory budgets, regular-grid compression tends to smooth, quantize, or remove sparse foreground structures. In contrast, our adaptive representation allocates its capacity according to the spatial distribution and complexity of the data, allowing it to preserve sparse and visually salient structures more effectively under the same storage budget. This advantage is particularly meaningful for highly sparse and spatially nonuniform volumes.}

\noindent\textbf{Time Comparison.}
\sm{To further analyze runtime cost, Appendix Table~\ref{tab:time_comparison} reports compression (training) time, decompression time, single-frame rendering time, and their total. The main cost of ESVR is incurred offline during optimization-based training, which can take up to several days when the largest volumes are trained sequentially on a single GPU. Once trained, however, ESVR requires no separate decompression stage: primitives are directly rasterized at query time, enabling millisecond-level rendering with low VRAM usage even for extremely large volumes. Traditional compressors encode much faster, but require full-volume decompression before rendering, after which the decompressed volume is visualized using DVR. Decompression time also grows significantly with data size, reaching tens to hundreds of seconds for datasets from Neuron onward. Furthermore, DVR runs out of memory on Woodbranch and Hemibrain. Therefore, ESVR is not intended as a fast one-time encoder, but instead offers a different trade-off: a higher offline preprocessing cost in exchange for a compact, directly renderable representation that supports interactive visualization in memory-limited settings where full-volume decompression or DVR becomes impractical.}
}

\sm{
\subsection{Comparison with VDB-based Methods}
\label{app:vdb}

\sm{
\noindent To better position ESVR with respect to sparse volumetric representations, we compare our method with OpenVDB~\cite{museth2013openvdb} and NanoVDB~\cite{museth2021nanovdb} in terms of compression efficiency, reconstruction quality, memory usage, and rendering performance. 
%
OpenVDB is an open-source implementation of the VDB sparse volume data structure~\cite{museth2013openvdb}. In this section, we use the term VDB-based representation to refer to sparse hierarchical grid representations following this data structure. OpenVDB denotes the CPU-oriented implementation used for conversion and storage, while NanoVDB converts the OpenVDB hierarchy into a compact GPU-friendly format optimized for traversal and rendering. 
We do not include NeuralVDB~\cite{kim2024neuralvdb}, which is a representative neural compression approach for VDB volumes, in our quantitative evaluation because there is currently no publicly available implementation suitable for reproducible evaluation.
Furthermore, NeuralVDB primarily focuses on neural compression of VDB volumes, whereas our evaluation targets directly renderable representations for which model size, reconstruction quality, rendering performance, and memory consumption can be measured consistently.

\begin{table}[!h]
\centering
\scriptsize
\setlength{\tabcolsep}{5pt}
\renewcommand{\arraystretch}{1.12}
\vspace{-0.5em}
\caption{\sm{Compression results on OpenVDB and NanoVDB under lossless/lossy setting. Lossy VDB results are generated under a Size-matched setting with ESVR.}}
\label{tab:vdb_nvdb_conversion}

\resizebox{\linewidth}{!}{
\begin{tabular}{ll|c|cc|cc}
\noalign{\hrule height 1.0pt}
\multirow{2}{*}{\textbf{Dataset}}
& \multirow{2}{*}{\textbf{Metric}}
& \multirow{2}{*}{\textbf{Ours}}
& \multicolumn{2}{c|}{\textbf{Lossless}}
& \multicolumn{2}{c}{\textbf{Lossy}} \\
\cline{4-7}
&
&
& \textbf{OpenVDB}
& \textbf{NanoVDB}
& \textbf{OpenVDB}
& \textbf{NanoVDB} \\
\hline

\multirow{2}{*}{Aneurysm}
& Size & 2.6MB & 0.7MB & 6.7MB & 0.7MB & 2.59MB \\
& PSNR & 51.54 & $\infty$ & $\infty$ & $\infty$ & 41.94 \\
\hline

\multirow{2}{*}{Breast}
& Size & 2.9MB & 2.8MB & 17.6MB & 2.8MB & 3.3MB \\
& PSNR & 46.00 & $\infty$ & $\infty$ & $\infty$ & 27.51 \\
\hline

\multirow{2}{*}{Neuron}
& Size & 2.8MB & 1.1GB & 1.8GB & 2.6MB & 3.1MB \\
& PSNR & 39.56 & $\infty$ & $\infty$ & 32.83 & 32.23 \\
\hline

\multirow{2}{*}{BigBrain}
& Size & 7.6MB & 1.8GB & 5.7GB & 7.8MB & 7.4MB \\
& PSNR & 21.53 & $\infty$ & $\infty$ & 7.54 & 7.52 \\
\hline

\multirow{2}{*}{Woodbranch}
& Size & 4.3MB & 18.8GB & 35.2GB & 4.8MB & 4.3MB \\
& PSNR & 40.60 & $\infty$ & $\infty$ & 28.43 & 28.41 \\
\hline

\multirow{2}{*}{\begin{tabular}[c]{@{}l@{}}Hemibrain\\(vDeltaA)\end{tabular}}
& Size & 4.2MB & 0.5GB & 2.2GB & 4.1MB & 4.2MB \\
& PSNR & 35.34 & $\infty$ & $\infty$ & 30.11 & 30.06 \\
\hline

\multirow{2}{*}{\begin{tabular}[c]{@{}l@{}}Hemibrain\\(lLN2P)\end{tabular}}
& Size & 5.8MB & 1.0GB & 4.5GB & 6.0MB & 6.4MB \\
& PSNR & 37.03 & $\infty$ & $\infty$ & 29.46 & 29.43 \\
\noalign{\hrule height 1.0pt}

\end{tabular}
}
\end{table}
\begin{figure}[tbh]
    \centering
    \includegraphics[width=0.9\linewidth]{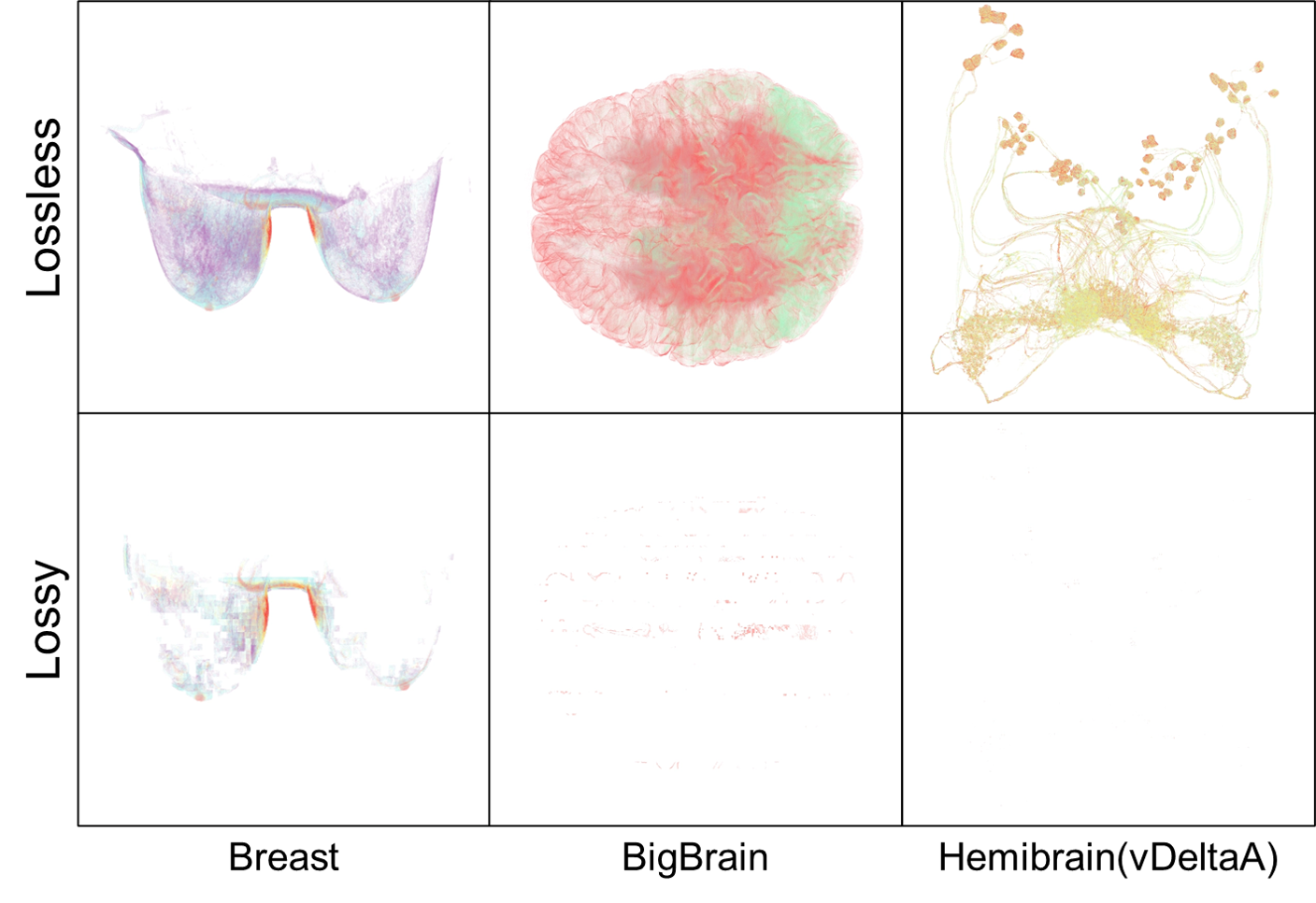}
    \caption{Rendering results of volumes converted into lossless/lossy VDB structure.}
    \label{fig:vdb-rendering}
\end{figure}

\noindent\textbf{Compression Results.}
Appendix Table~\ref{tab:vdb_nvdb_conversion} summarizes the results on both lossless OpenVDB/NanoVDB conversion and lossy fixed-rate VDB compression following Zellmann et al.~\cite{zellmann2025gpuvolumerenderinghierarchical}. 
In the lossless setting, OpenVDB and NanoVDB preserve the original voxel values exactly, resulting in infinite PSNR. However, their storage requirements remain substantially larger than those of ESVR. For example, the lossless NanoVDB representation requires 5.7 GB for BigBrain and 35.2 GB for Woodbranch, whereas ESVR represents the same datasets using only 7.6 MB and 4.3 MB, respectively. Although VDB structures efficiently remove empty regions, they remain fundamentally grid-based representations and therefore must preserve voxel information at the original grid resolution.

\noindent We further evaluate lossy VDB compression under a size-matched setting with ESVR. While lossy OpenVDB/NanoVDB can achieve a similar storage size, reconstruction quality degrades substantially. For example, on Breast, ESVR achieves 46.00 dB PSNR at 2.9 MB, whereas lossy NanoVDB achieves only 27.51 dB at 3.3 MB. Similar trends can be observed across all datasets. These results indicate that matching the compact representation size of ESVR with a grid-based VDB representation requires aggressive voxel removal and quantization.
The visual impact of this compression is shown in Appendix Figure~\ref{fig:vdb-rendering}. Although some datasets exhibit only moderate PSNR degradation numerically, the rendered results reveal severe structural loss. In particular, many thin and sparse structures disappear entirely, leaving large empty regions in the rendered volume. This behavior becomes increasingly pronounced for high-resolution datasets because VDB compression ultimately operates by deciding which voxel blocks should be retained or discarded. As volume resolution increases, preserving fine structures under a strict storage budget becomes progressively more difficult.

\begin{table}[!h]
\centering
\scriptsize
\renewcommand{\arraystretch}{1.12}
\caption{\sm{Rendering performance and VRAM usage on lossless/lossy compression using NanoVDB.}}
\label{tab:nanovdb_rendering}
\resizebox{0.7\linewidth}{!}{%
\begin{tabular}{ll|c|cc}
\noalign{\hrule height 1.0pt}
\textbf{Dataset} 
& \textbf{Metric} 
& \textbf{Ours} 
& \textbf{Lossless} 
& \textbf{Lossy} \\
\hline

\multirow{2}{*}{Aneurysm}
& FPS  & 223.19 & 204.55 & 239.28 \\
& VRAM & 1.64GB & 1.56GB & 1.55GB \\
\hline

\multirow{2}{*}{Breast}
& FPS  & 213.49 & 145.75 & 257.77 \\
& VRAM & 1.64GB & 1.57GB & 1.55GB \\
\hline

\multirow{2}{*}{Neuron}
& FPS  & 131.08 & 32.32 & 387.15 \\
& VRAM & 1.64GB & 3.39GB & 1.56GB \\
\hline

\multirow{2}{*}{BigBrain}
& FPS  & 48.64 & 10.76 & 372.18 \\
& VRAM & 1.74GB & 7.29GB & 1.56GB \\
\hline

\multirow{2}{*}{Woodbranch}
& FPS  & 43.42 & \textsc{oom} & 425.28 \\
& VRAM & 1.71GB & \textsc{oom} & 1.56GB \\
\hline

\multirow{2}{*}{\begin{tabular}[c]{@{}l@{}}Hemibrain\\(vDeltaA)\end{tabular}}
& FPS  & 137.95 & 90.31 & 462.94 \\
& VRAM & 1.69GB & 3.75GB & 1.56GB \\
\hline

\multirow{2}{*}{\begin{tabular}[c]{@{}l@{}}Hemibrain\\(lLN2P)\end{tabular}}
& FPS  & 86.76 & 69.15 & 469.26 \\
& VRAM & 1.67GB & 6.07GB & 1.56GB \\

\noalign{\hrule height 1.0pt}
\end{tabular}
}
\end{table}
\noindent\textbf{Rendering Performance.} 
Appendix Table~\ref{tab:nanovdb_rendering} reports NanoVDB rendering performance. 
For rendering, we implement a simple OptiX-based HDDA ray-marching renderer for NanoVDB and report FPS and peak VRAM usage. We omit OpenVDB rendering performance because its CPU-oriented traversal model is not directly comparable to our GPU-based renderer, and instead use NanoVDB as the relevant GPU-based baseline for interactive rendering.
Lossless NanoVDB generally renders more slowly than ESVR while requiring substantially higher memory usage. In contrast, lossy NanoVDB often achieves higher FPS than ESVR. However, this speedup must be interpreted together with the quality degradation shown in Appendix Table~\ref{tab:vdb_nvdb_conversion} and Appendix Figure~\ref{fig:vdb-rendering}. Because aggressive compression removes a large fraction of the original voxel data, the renderer traverses significantly fewer active regions, resulting in higher rendering performance at the expense of structural fidelity.

\noindent These observations highlight a fundamental difference between the two representations. VDB-based methods remain resolution-dependent sparse grids whose memory and compression behavior are tied to the underlying voxel discretization. Consequently, their effectiveness decreases as volume resolution increases or as the volume becomes less sparse. This trend is visible in Woodbranch, where the denser occupancy leads to noticeably weaker compression compared to the highly sparse Hemibrain datasets.
In contrast, ESVR is a primitive-based representation. Instead of preserving individual voxels, it approximates volumetric structures using a compact set of anisotropic ellipsoidal primitives. As a result, its storage complexity and rendering performance depend primarily on the number of primitives rather than the original grid resolution. This enables ESVR to maintain compact model sizes while preserving sparse anisotropic structures and supporting interactive rendering. Overall, ESVR provides a more balanced trade-off between storage size, reconstruction fidelity, memory consumption, and rendering performance for extremely large sparse volumetric datasets.
}
}

\section{Ablation}



\subsection{Primitive Type Comparison}
\label{app:syn_primitive}
\begin{figure}[!h]
 \centering
 \includegraphics[width=\linewidth]{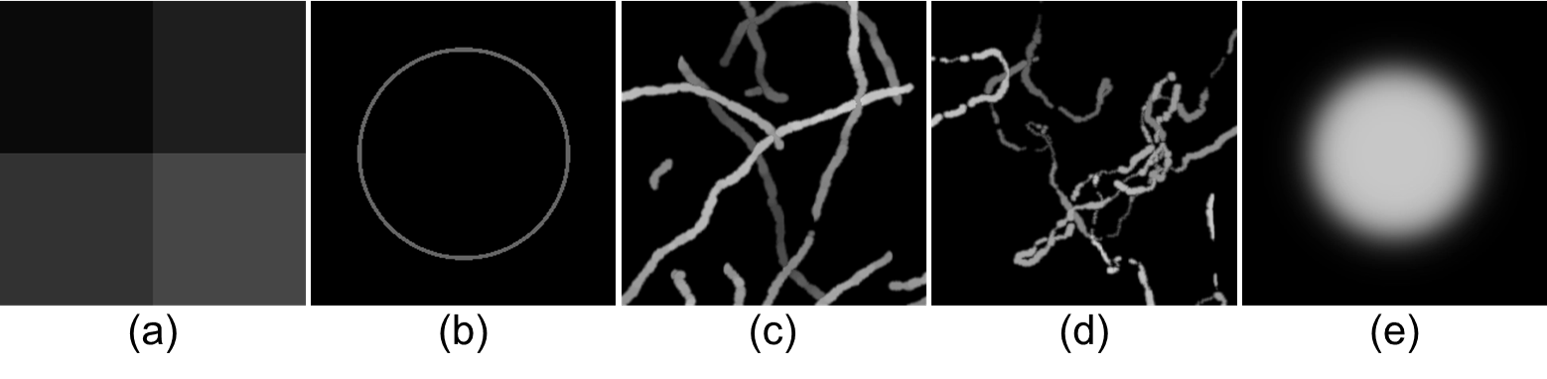}
   \caption{
   2D slice of synthetic datasets used for primitive type comparison. (a) 8area, (b) EmptySphere, (c) SynthTube, (d) SynthWiggle, \sm{(e) DiffuseScalar}
   }
 \label{fig:synthetic_data}
\end{figure}
\noindent To further evaluate the expressiveness of our differentiable ellipsoids, we construct 
five synthetic datasets, as shown in Appendix Figure~\ref{fig:synthetic_data}, and compare them with Gaussian primitives in terms of FPS, PSNR, and average primitive hits per voxel. 
Appendix Figure~\ref{fig:synthetic_data} (a) shows a dense volume divided into eight chunks (8area), while Appendix Figure~\ref{fig:synthetic_data} (b) represents an empty sphere with only a surface (EmptySphere). Appendix Figures ~\ref{fig:synthetic_data} (c) and (d) correspond to tube-like structures (SynthTube and SynthWiggle) \sm{and (e) sphere with diffuse boundaries (DiffuseScalar)}.
As shown in Appendix Table~\ref{tab:syn_primitive}, our method achieves lower PSNR than Gaussians in dense data. However, it consistently provides significantly higher FPS and fewer average primitive hits per voxel, demonstrating improved computational efficiency.
In contrast, results on EmptySphere show that our differentiable ellipsoids better capture surface structures than Gaussians. Furthermore, on SynthTube and SynthWiggle, our method achieves both higher PSNR and faster rendering speed, indicating strong expressiveness for sparse and anisotropic structures.
\sm{Notably, DiffuseScalar's smooth boundary blending with Gaussian tails achieves superior reconstruction quality compared to ellipsoid overlaps. While our ellipsoid representation demonstrates advantages in FPS and primitive hits as intended, it exhibits fundamental limitations in expressiveness for representing smooth boundaries.} 

\begin{table}[!htb]
\caption{
Ablation study on primitive types (Ellipsoid vs.\ Gaussian) before and after pruning on synthetic datasets. Sparsity improvement is measured through average hits per voxel and primitive count reduction. Avg PrimHits denotes the average number of primitives per voxel, computed over voxels with non-zero primitive hits.}
\centering
\small
\renewcommand{\arraystretch}{1.1}
\resizebox{\linewidth}{!}{
\begin{tabular}{l|l|cc|cc}
\noalign{\hrule height 1.0pt}
    & & \multicolumn{2}{c|}{\textbf{Ellipsoid}} & \multicolumn{2}{c}{\textbf{Gaussian}} \\
    \textbf{Dataset} & \textbf{Metric} & \textit{Before} & \textit{After} & \textit{Before} & \textit{After} \\
\hline
\multirow{4}{*}{8area} 
        & Size (MB) & 4.18 (97k) & 1.27 (29k) & 4.25 (99k) & 1.27 (29k) \\
        & 3D PSNR (dB) & 41.73 & 42.67 & 45.86 & 47.93 \\
        & FPS & 59.98 & 60.12 & 3.60 & 6.32 \\
        & Avg PrimHits & 7.84 & 5.68 & 15.44 & 9.97  \\
\hline
\multirow{4}{*}{EmptySphere} 
        & Size (MB) & 6.66 (155k) & 2.26 (53k) & 6.55 (152k) & 2.23 (52k) \\
        & 3D PSNR (dB) & 46.13 & 47.40 & 38.10 & 43.91 \\
        & FPS & 167.33 & 291.26 & 9.25 & 18.90  \\
        & Avg PrimHits & 9.79 & 3.53 & 26.47 & 13.50  \\
\hline
\multirow{4}{*}{SynthTube} 
        & Size (MB) & 2.61 (61k) & 0.89 (21k) & 2.74 (64k) & 0.93 (22k) \\
        & 3D PSNR (dB) & 71.51 & 66.25 & 63.02 & 64.66  \\
        & FPS & 183.78 & 268.57 & 19.08 & 50.26  \\
        & Avg PrimHits & 14.99 & 9.29 & 34.69 & 19.26 \\
\hline
\multirow{4}{*}{SynthWiggle} 
        & Size (MB) & 1.42 (33k) & 0.48 (11k) & 1.45 (34k) & 0.49 (11k) \\
        & 3D PSNR (dB) & 73.46 & 67.88 & 65.18 & 65.18  \\
        & FPS & 226.76 & 329.07 & 31.79 & 88.39  \\
        & Avg PrimHits & 12.56    & 7.31 & 24.89 & 12.28  \\
\hline
\multirow{4}{*}{\sm{DiffuseScalar}} 
        & Size (MB) & 4.54 (105K) & 1.35 (29k) & 4.35 (106k) & 1.36 (31k) \\
        & 3D PSNR (dB) & 48.90 & 48.18 & 52.28 & 53.84  \\
        & FPS & 124.63 & 157.74 & 5.04 & 12.76  \\
        & Avg PrimHits & 5.71 & 5.01 & 13.79 & 8.25 \\
\hline
\noalign{\hrule height 1.0pt}
\end{tabular}
}
\label{tab:syn_primitive}
\end{table}

\subsection{Pruning strategy}
We additionally provide qualitative results in Appendix Figure~\ref{fig:importance_score} to support the pruning strategy.
As shown in the figure, FA-based scoring better preserves fine vascular structures and high-frequency details, while intensity and scale cues help maintain dense regions. Ellipsoids with higher FA scores tend to cluster around structurally informative regions, demonstrating that FA effectively captures structural importance.
These observations are consistent with the quantitative results shown in Table \ref{tab:loss_ablation} in the main text, where combining FA with intensity and scale cues yields the best reconstruction quality.

\begin{figure}[!th]
 \centering
 \includegraphics[width=1.0\linewidth]{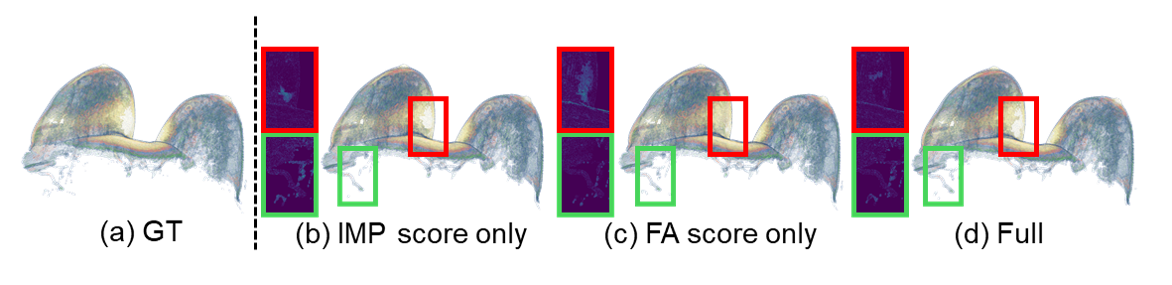}
   \caption{
    Qualitative results highlighting the effect of the FA score on rendering quality of Breast dataset. 
   }
 \label{fig:importance_score}
\end{figure}

\subsection{Learnable K}
\label{app:learnable-k}
While we use $k$ as a fixed hyperparameter in the main setting, we additionally evaluate a learnable $k$ per ellipsoid, as well as a smoother configuration with $k=3$ in Appendix Table \ref{app:k_ablation}.
We observe that the effectiveness of learnable $k$ depends on the dataset. In some cases (e.g., Aneurysm and BigBrain), a fixed $k$ achieves higher PSNR than the learnable variant, as the learned values tend to saturate around $k\approx5$. In contrast, for datasets with denser regions such as Woodbranch, the learnable $k$ yields the highest PSNR.
However, the PSNR improvement is marginal, while rendering performance degrades due to the need to compute adaptive cutoff regions for each primitive. This introduces a trade-off between PSNR and FPS depending on the dataset. Given the relatively small PSNR gains compared to the significant drop in FPS, we adopt a fixed $k$ as a hyperparameter in our final design.
%

\begin{table}[!htb]
\centering
\small
\setlength{\tabcolsep}{4pt}
\renewcommand{\arraystretch}{1.1}
\caption{Comparison of different sharpness ($k$) settings of differentiable ellipsoids. Each cell reports 3D PSNR (dB) / number of primitives / FPS. For BigBrain and Woodbranch, we use one of the chunks for evaluation.}
\begin{tabular}{l|ccc}
\noalign{\hrule height 1.0pt}
\textbf{Dataset} & \textbf{k=5} & \textbf{k=3} & \textbf{Learnable k} \\
\hline

Aneurysm 
& \textbf{50.96} / 33.2k / \textbf{250.85 }
& 48.73 / 32.7k / 196.10 
& 49.76 / 33.8k / 182.12 \\


BigBrain
& \textbf{18.65} / 47.3k / \textbf{153.38 }
& 18.54 / 42.6k / 142.71 
& 18.52 / 43.8k / 132.51 \\

Woodbranch
& 41.02 / 11.9k / \textbf{238.70 }
& 41.30 / 12.7k / 235.18 
& \textbf{41.33} / 12.5k / 225.31 \\

\noalign{\hrule height 1.0pt}
\end{tabular}
\label{app:k_ablation}
\end{table}

\subsection{Boundary Sharpness}
\label{app:boundary-sharpness}


We evaluate the effect of boundary sharpness controlled by parameter $k$ in Appendix Figure~\ref{fig:ellipsoid_k}.
Raw evaluation metrics ($\textit{L}_1$ loss, DSSIM loss, and PSNR) are recorded every 100 iterations. Since the loss is computed on subvolumes, directly plotting the raw loss and PSNR values does not accurately reflect convergence and may appear to fluctuate. To improve visual clarity, a moving average filter is applied to the raw logs. Specifically, a window size of $n = 10$ is used for Appendix Figure~\ref{fig:ellipsoid_k}.
As shown in  Appendix Figure~\ref{fig:ellipsoid_k}, soft boundaries $k=5$ yield stable convergence and high PSNR, while increasing $k$ progressively slows and destabilizes optimization as gradients become increasingly localized near sharp boundaries. At $k=5000$, PSNR saturates immediately and fails to improve, confirming that hard boundaries suppress the gradient signal needed for effective parameter updates. This behavior contrasts with EVER, where hard ellipsoids remain trainable through analytic transmittance integration. In our volumetric setting, near-zero spatial gradients outside the ellipsoid footprint make hard boundaries unsuitable. 


\begin{figure}[htb]
 \centering
 \includegraphics[width=0.9\linewidth]{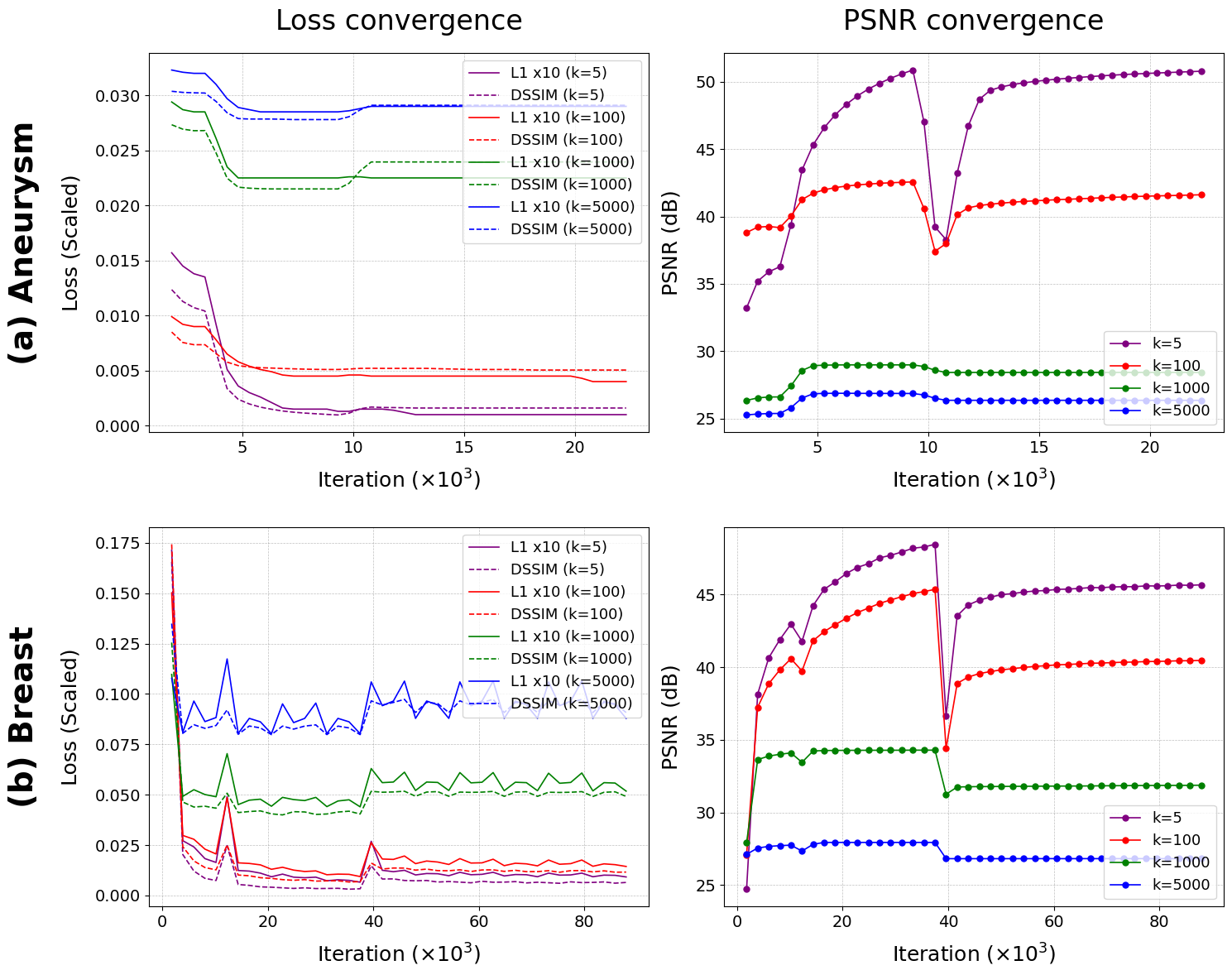}
   \caption{Ablation on boundary sharpness (k) of differentiable ellipsoids on Aneurysm and Breast datasets.}
 \label{fig:ellipsoid_k}
\end{figure}

\subsection{\sm{Effect of Depth Bins}}
\sm{
\noindent We investigate the impact of the number of depth bins used in the per-primitive ray sampling renderer. Appendix Figure~\ref{fig:depth-bin} reports rendering performance (FPS), GPU memory usage, and rendering quality for 1, 4, 8, and 16 depth bins.
Using multiple depth bins generally improves rendering performance by increasing intra-ray parallelism. However, the benefit saturates when the number of bins increases too much. Across all tested datasets, 4 and 8 bins achieve the highest rendering performance, whereas 16 bins consistently reduces FPS and can even perform worse than the single-bin configuration due to the additional overhead of maintaining and compositing per-bin intermediate buffers.
GPU memory consumption exhibits a similar trend. The 4-bin configuration requires the least VRAM across all datasets, while the 16-bin configuration increases memory usage because of the larger amount of intermediate storage required for per-bin accumulation and compositing.
\noindent To assess rendering quality, we use the single-bin renderer as a reference and report the relative mean squared error (MSE). The relative MSE remains low for all tested configurations, indicating that depth binning introduces only minor image-space differences. No monotonic relationship is observed between the number of bins and rendering quality. Instead, the optimal bin count varies across datasets because changing the number of bins alters the placement of depth-bin boundaries along each ray, which can slightly affect sampling and alpha compositing near those boundaries. Consequently, the observed differences depend more strongly on primitive distributions, overlap patterns, and dataset characteristics than on the bin count itself.
Based on these results, we use 8 depth bins as the default configuration throughout the paper, providing a robust balance between rendering performance, memory consumption, and image quality. For memory-constrained scenarios, 4 bins offer a more memory-efficient alternative while maintaining comparable visual fidelity.
\begin{figure}[h]
    \centering
    \includegraphics[width=1.0\linewidth]{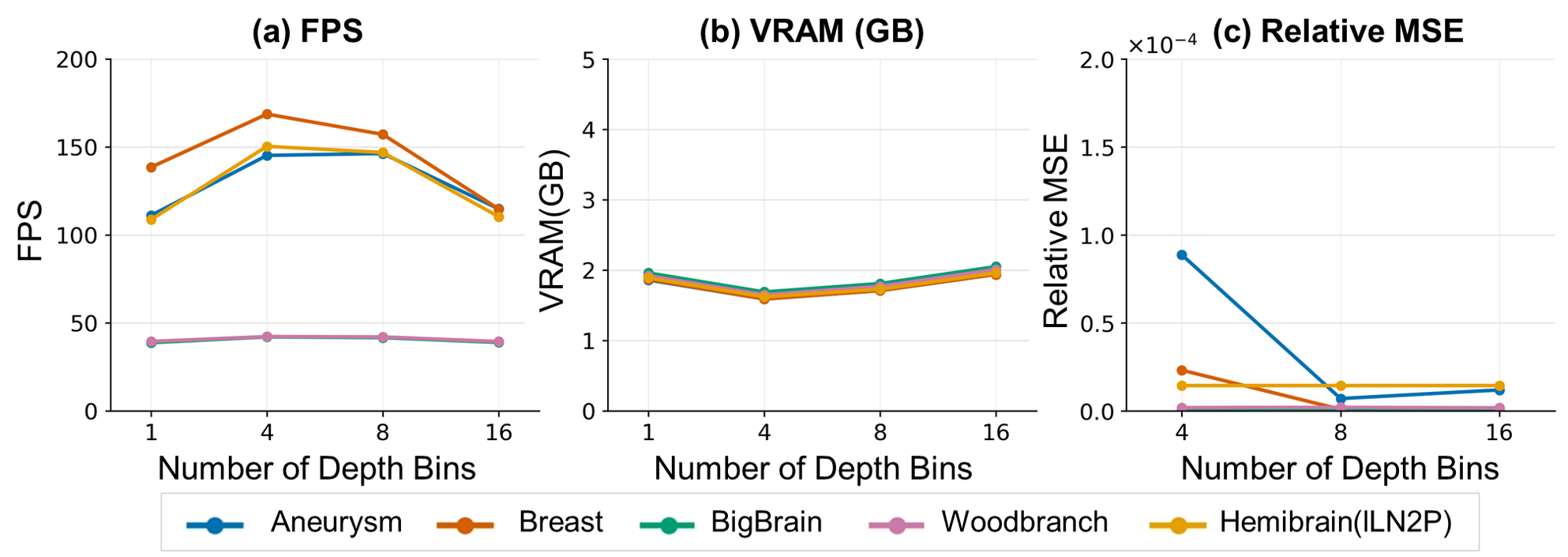}
    \caption{\sm{Effect of depth bin count on (a) rendering performance (FPS), (b) GPU memory usage, and (c) rendering quality.}}
    \label{fig:depth-bin}
\end{figure}
}







\end{document}